\documentclass[
    amsmath,
    amssymb,
    amsfonts,
    prx,
    aps,
    floatfix,
    twocolumn,
    longbibliography,
    superscriptaddress,
]{revtex4-2}

\usepackage[utf8]{inputenc}
\usepackage[T1]{fontenc}
\usepackage{physics}
\usepackage{bbm, bm}
\usepackage{upgreek}
\usepackage{nicefrac}

\usepackage{graphicx}
\usepackage[dvipsnames]{xcolor}

\usepackage[breaklinks=true, colorlinks]{hyperref}
\hypersetup{allcolors=blue}
\usepackage[capitalize, noabbrev]{cleveref}

\usepackage{booktabs}
\usepackage{makecell}
\usepackage{multirow}
\usepackage{cellspace}
\newcommand{\rr}{{\vb{r}}}
\newcommand{\RR}{{\vb{R}}}

\newcommand{\xc}{{\text{xc}}}

\newcommand{\fa}{{\text{FA}}}
\newcommand{\ms}{{\text{MS}}}
\newcommand{\ext}{{\text{ext}}}
\newcommand{\oep}{{\text{OEP}}}
\newcommand{\iks}{{\text{IKS}}}
\newcommand{\exx}{{\text{EXX}}}

\DeclareMathOperator*{\argmin}{argmin}
\DeclareMathOperator*{\argmax}{argmax}

\crefname{equation}{Equation}{Equations}
\crefname{figure}{Figure}{Figures}
\crefname{table}{Table}{Tables}
\creflabelformat{equation}{#2(#1)#3}
\creflabelformat{figure}{#2(#1)#3}
\creflabelformat{table}{#2(#1)#3}

\makeatletter
\newcommand{\setsref}[2]{%
    \expandafter\newcommand\csname sref@#1\endcsname{#2}%
}
\newcommand{\sref}[1]{%
    \@ifundefined{sref@#1}{%
        \PackageError{sref}{Unknown supplementary label '#1'}%
        {Add a mapping with \string\setsref\space in the main.tex preamble.}%
    }{\sref@format#1\@nil}%
}
\def\sref@format#1:#2\@nil{%
    \@ifundefined{sref@type@#1}{%
        \PackageError{sref}{Unknown supplementary prefix '#1'}%
        {Use sec, eq, fig or tab as the label prefix.}%
    }{Supplementary~\csname sref@type@#1\endcsname{\csname sref@#1:#2\endcsname}}%
}
\newcommand{\sref@type@sec}[1]{Note~#1}
\newcommand{\sref@type@eq}[1]{Equation~#1}
\newcommand{\sref@type@fig}[1]{Figure~#1}
\newcommand{\sref@type@tab}[1]{Table~#1}
\makeatother
\setsref{sec:gradients}{1}
\setsref{sec:regularization}{2}
\setsref{sec:params}{3}
\setsref{sec:comparison}{4}
\setsref{sec:hyperparams}{5}
\setsref{sec:details}{6}
\setsref{tab:method-comparison}{1}
\setsref{tab:hyperparams}{2}
\setsref{tab:rydberg-basis}{3}
\setsref{tab:energy-errs}{4}
\setsref{fig:comparison}{1}
\setsref{fig:ablation-metrics}{2}
\setsref{fig:ablation-profiles}{3}
\setsref{fig:potential-gallery}{4}
\setsref{eq:exponent-params}{45}

\begin{document}

\title{Multipole splats for optimized and inverted effective potentials}

\author{Matija Medvidović}
\email{mmedvidovic@ethz.ch}
\affiliation{Institute for Theoretical Physics, ETH Zürich, 8093 Zürich, Switzerland}

\author{Angel Rubio}
\affiliation{Max Planck Institute for the Structure and Dynamics of Matter, Luruper Chaussee 149, 22761 Hamburg, Germany}
\affiliation{Center for Computational Quantum Physics, Flatiron Institute, 162 5th Avenue, New York, NY 10010, USA}
\affiliation{Initiative for Computational Catalysis, Flatiron Institute, 162 5th Avenue, New York, NY 10010, USA}

\author{Juan Carrasquilla}
\affiliation{Institute for Theoretical Physics, ETH Zürich, 8093 Zürich, Switzerland}

\date{September 22, 2026}

\begin{abstract}
   In modern density functional theory, effective Hamiltonians are constructed to reproduce densities and forces of electrons at equilibrium. However, their nonlocal potentials leave systematic errors in spectral properties and real-time dynamics. Access to local optimized or inverted effective potentials would remove this limitation, but the numerical fragility in finite orbital bases has long prevented their wide adoption. Here, we introduce \textit{multipole splats}, a class of trial potentials that carry the correct asymptotic decay required to support the unoccupied spectrum. By connecting the computation of effective potentials to variational and supervised variants of Hamiltonian learning, we recast both problems as stable nonlinear optimization formulated directly in standard orbital basis sets and applicable to any hybrid functional approximation. The resulting solver allows us to resolve spatial profiles of exchange-correlation potential errors during molecular dissociation and accurately reconstruct key excited states without empirical asymptotic corrections. We also show the deviation from the ionization potential theorem for different exchange-correlation approximations on a dataset of molecular systems. Multipole splats extract insights from established approximations and provide capacity for robust dataset generation for downstream processing and learning.
\end{abstract}

\maketitle

\begin{figure*}[t]
    \centering
    \includegraphics[width=\linewidth]{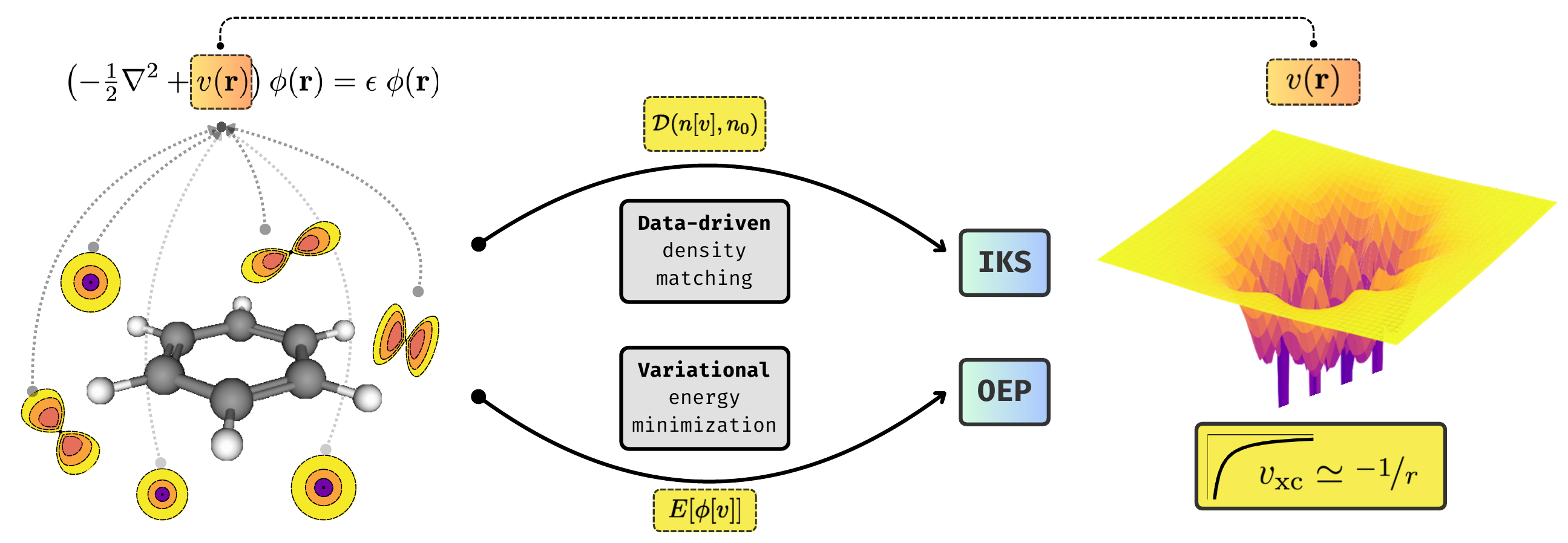}
    \caption{
        \textbf{A schematic of the multipole splat potential optimization.}
        Two optimization modes are shown. Variational optimization estimates optimized effective potentials by minimizing the total energy under a given exchange-correlation approximation while the supervised optimization estimates inverted Kohn-Sham potentials by minimizing the quantum relative entropy between the target density and the density generated by the trial potential. Floating Gaussian multipoles (\textit{splats}) parameterize the trial potential.
    }
    \label{fig:diagram}
\end{figure*}

\section*{Introduction}
\label{sec:introduction}

Density functional theory (DFT) has offered controlled approximations of many-electron phenomena in molecular and condensed matter systems for decades. Large parts of downstream computational science see DFT as the only interface with quantum phenomena. Materials science~\cite{curtaroloHighthroughputHighwayComputational2013}, quantum chemistry~\cite{huangCentralRoleDensity2023, martinElectronicStructureBasic2020}, and biology~\cite{vanmourikDensityFunctionalTheory2014} all rely on the underlying effective single-electron description~\cite{hohenbergInhomogeneousElectronGas1964, kohnSelfConsistentEquationsIncluding1965} supplied by DFT solvers. This role as a methodological bottleneck has been tightened by the recent developments in data-driven science supported by artificial intelligence (AI), where DFT is used as a data factory for models of quantum matter~\cite{unkeMachineLearningForce2021, jacobsPracticalGuideMachine2025a}.

While exact in theory, the Kohn-Sham (KS)~\cite{kohnSelfConsistentEquationsIncluding1965} approach suffers from well-catalogued limitations of approximate exchange-correlation (XC) functionals and finite basis sets \cite{martinElectronicStructureBasic2020, linMathematicalIntroductionElectronic2019, beckePerspectiveFiftyYears2014}. To mitigate self-interaction errors and model thermochemistry, modern generalized Kohn-Sham (GKS) functionals~\cite{perdewRationaleMixingExact1996, kummelOrbitaldependentDensityFunctionals2008a} mix in fractions of orbital-dependent exact exchange. The local multiplicative KS potential is replaced by a nonlocal operator, hiding real-space force fields, spectral lines, and the guarantees of KS theorems.

Optimized effective potentials (OEP)~\cite{kummelOptimizedEffectivePotential2003} restore the true independent-electron physics based on local potentials. The OEP is the multiplicative local potential whose orbitals minimize the total variational GKS energy. In the inverse Kohn-Sham (IKS) problem \cite{shiInverseKohnSham2021}, a local potential is recovered from target density data. In the original formulation~\cite{sharpVariationalApproachUnipotential1953, talmanOptimizedEffectiveAtomic1976}, evaluating the OEP required inverting the KS response function, which is strongly ill-conditioned in finite basis sets because small response eigenvalues amplify numerical noise and can generate oscillatory, unphysical potentials~\cite{staroverovOptimizedEffectivePotentials2006, heaton-burgessOptimizedEffectivePotentials2007, trushinNumericallyStableOptimized2021}. Regardless, potentials are most commonly represented as linear combinations of auxiliary basis functions \cite{yangDirectMethodOptimized2002, wuAlgebraicEquationIterative2003, heaton-burgessOptimizedEffectivePotentials2007, heaton-burgessOptimizedEffectivePotentials2008, bulatOptimizedEffectivePotentials2007, hesselmannNumericallyStableOptimized2007, kollmarOptimizedEffectivePotential2007, jacobUnambiguousOptimizationEffective2011, trushinNumericallyStableOptimized2021, trushinAvoidingSpinContamination2023, trushinImprovingExchangeCorrelationPotentials2025} or replaced by semi-analytical approximations~\cite{kriegerConstructionApplicationAccurate1992, kriegerSystematicApproximationsOptimized1992, kummelOptimizedEffectivePotential2003, kummelSimpleIterativeConstruction2003a, beckeSimpleEffectivePotential2006a}. The data-driven IKS uses physical objectives like the Wu-Yang (WY)~\cite{wuDirectOptimizationMethod2003} or the Zhao-Morrison-Parr~\cite{zhaoConstrainedsearchMethodDetermine1993} functionals but faces the same finite-basis expressivity problem.

The OEP and IKS technical debt is still growing, as accurate potential data is needed at larger scales. Precise IKS potentials are natural targets for machine-learned XC functionals \cite{vossMachineLearningAccuracy2024} and force fields \cite{unkeMachineLearningForce2021} through the Hellmann-Feynman theorem~\cite{linMathematicalIntroductionElectronic2019}. Modern calculations with neural quantum states (NQS) \cite{hermannDeepneuralnetworkSolutionElectronic2020, pfauInitioSolutionManyelectron2020, medvidovicNeuralnetworkQuantumStates2024} produce precise densities uniquely positioned to inform XC functional design through inversion. Reference data errors propagate into machine-learned XC functionals and force fields, affecting downstream predictions.

Time-dependent DFT (TD-DFT)~\cite{rungeDensityFunctionalTheoryTimeDependent1984, marquesTimeDependentDensityFunctional2006, ullrichSnapshotTimedependentDensityfunctional2025} inherits the same problem through its sensitivity to the ground-state potential. Coulomb-like potential tails are required to support key excited states~\cite{dreuwFailureTimeDependentDensity2004, maitraChargeTransferTimeDependentDensity2017, kaurWhatAccuracyLimit2019, trushinImprovingExchangeCorrelationPotentials2025}. Beyond the linear response regime, numerical problems are further amplified~\cite{mundtOptimizedEffectivePotential2006}. In both the data-driven and spectral settings, solver accuracy is a real limitation.

We bypass this inversion by unifying OEP and IKS as Hamiltonian learning problems over parameterized trial potentials. Calculations are reframed as explicitly nonlinear optimization tasks, with a variational energy loss in the OEP case and a quantum density-matching loss in the IKS case, as shown on \cref{fig:diagram}. The KS eigensolver and first-order perturbation theory are connected into a differentiable solver enabling end-to-end optimization. Practical numerical stability is achieved by avoiding ill-conditioned inversions and supplemented by a physics-informed regularizer designed to select a smooth potential.

Trial potentials are built from \textit{multipole splats}, a superposition of floating Gaussian monopoles and dipoles with optimizable charges, positions, and widths. The potential is modeled as a Coulomb field of a fictitious charge distribution, making it a natural solver output and exactly guaranteeing asymptotic decay $v_\xc \simeq \nicefrac{-\gamma}{r}$ at all times.
Asymptotics are built in as a natural constraint rather than left to emerge. Multipole splats inherit tools from Gaussian splats \cite{kerbl3DGaussianSplatting2023}, a now-standard method in computer vision for complex three-dimensional scene reconstruction as clouds of floating Gaussians optimized through their projections (splats). We generalize splatting from pixel space to the atomic orbital basis, exploiting standard Gaussian integral technology and enabling reuse of standard quantum chemistry integral routines. Gaussian splats have recently been used to predict electronic densities from data~\cite{elsborgELECTRACartesianNetwork2026, elsborgGlobalPlaneWaves2026}.

The WY functional~\cite{wuDirectOptimizationMethod2003, shiInverseKohnSham2021} is a standard objective for IKS inversion. To establish a connection between IKS and statistical learning theory, we show that maximizing the WY functional corresponds to relative entropy minimization at zero temperature. The Kohn-Sham state belongs to a quantum \emph{exponential family}~\cite{wainwrightGraphicalModelsExponential2008, hasegawaExponentialMixtureFamilies1997}, with the potential as its natural parameter and the density as its sufficient statistic. In that formalism, minimizing the quantum relative entropy over admissible potentials is a maximum-likelihood~\cite{jaynesInformationTheoryStatistical1957, anshuSampleefficientLearningInteracting2021} fit that matches the KS density to the target density. Retaining finite $\beta$ gives the finite-temperature objective derived in Methods.

Here, we show real-space signatures of modern hybrid XC approximations for molecular systems, using multipole splats.
We resolve XC potential errors in molecular dissociation by comparing OEPs against IKS reference potentials from coupled-cluster calculations with singles, doubles and perturbative triples (CCSD(T)). These comparisons probe one-electron cancellation, asymmetric dissociation and bond breaking, separating molecular-region errors from asymptotic behavior. We also recover OEP KS gaps and highest occupied molecular orbital (HOMO) eigenvalues across a dataset of neutral molecular systems, demonstrating access to Rydberg states supported by the $\nicefrac{-1}{r}$ exact-exchange (EXX) multipole splat potential without empirical asymptotic corrections. The OEP gap is systematically below the GKS gap, exposing the spectral redistribution associated with localizing the potential. Finally, we compute OEP exchange source densities for larger $\pi$-conjugated molecules, directly mapping out electronic delocalization at the scale of $\sim 70$ electrons. These results reveal real-space fingerprints of modern exchange-correlation tradeoffs without inverting the KS response.

\section*{Results}

\subsection*{Multipole splats and direct optimization}

The effective potential is the central quantity in Kohn-Sham~\cite{hohenbergInhomogeneousElectronGas1964, kohnSelfConsistentEquationsIncluding1965} density functional theory. For $N$ interacting electrons, a new non-interacting system is constructed such that the two electronic densities $n$ match. Surrogate KS electrons are described by KS orbitals $\phi = \{ \phi _1 , \ldots, \phi _N \}$. Therefore, we need to find a single-particle potential $v$ such that the orbitals that solve
\begin{equation}
\label{eq:ks-multiplicative}
    \left( -\tfrac12 \laplacian + v(\rr) \right) \phi _k (\rr) = \epsilon _k \phi _k (\rr)
\end{equation}
minimize the total energy functional $E[\phi]$~\cite{martinElectronicStructureBasic2020}.

To reinterpret orbital-dependent GKS functionals in the KS language, we need computationally efficient access to optimized effective potentials for orbital-dependent functionals. The OEP is defined as the multiplicative local potential yielding orbitals (via \cref{eq:ks-multiplicative}) with minimal energy
\begin{equation}
\label{eq:oep-variational}
    v_\oep = \argmin _v E _\text{GKS} [\phi[v]] \; ,
\end{equation}
for a given GKS total energy functional $E_\text{GKS}$.

Alternatively, the exact KS potential can be reconstructed from the exact electronic density $n_0$ in a data-driven way, bypassing human-designed XC functional approximations. In the inverse Kohn-Sham formalism~\cite{shiInverseKohnSham2021}, we construct the potential through density matching,
\begin{equation}
\label{eq:iks-data}
    v_\iks = \argmin _v \mathcal{D} [n [v], n_0] \; ,
\end{equation}
where $n[v]$ is the density built from the solutions of \cref{eq:ks-multiplicative} and $\mathcal{D}[\cdot, \cdot]$ is a measure of functional divergence between normalized densities.

The IKS potential and the OEP share a common language in \cref{eq:oep-variational,eq:iks-data}. We propose a unified approach to effective KS potential optimization in IKS and OEP frameworks. We view IKS as a data-driven density matching problem and the OEP as a variational energy optimization problem, as outlined in \cref{fig:diagram}. Access to precise OEPs and IKS restores a local KS representation while maintaining the accuracy of higher levels of theory.

\begin{figure*}[t]
    \centering
    \includegraphics[width=\linewidth]{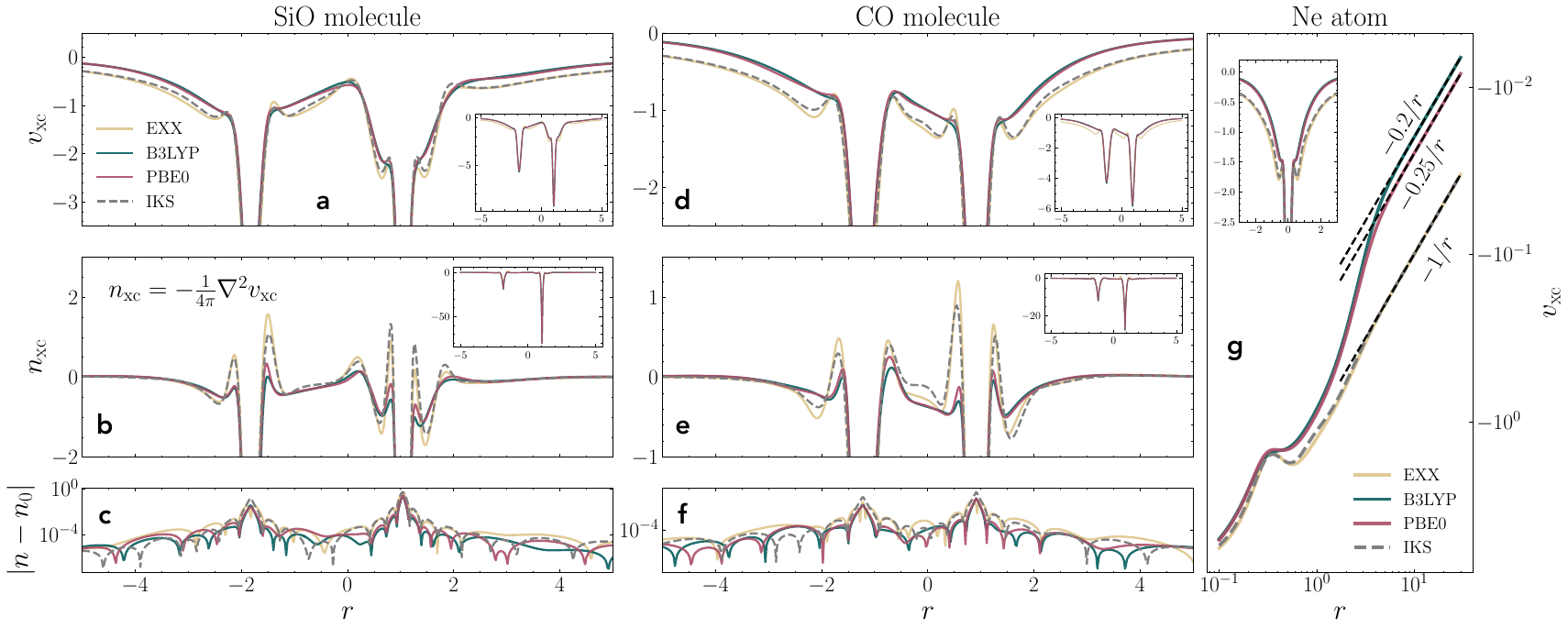}
    \caption{
        \textbf{Spatial profiles of effective potentials and their source densities.}
        All quantities given in Hartree atomic units.
        \textbf{a,b}:
        The OEP and IKS potentials and their XC source densities for the silicon monoxide (SiO) molecule using the EXX, B3LYP and PBE0 functionals as well as the IKS with respect to the independently computed CCSD(T) reference density. The orbitals were represented using the \texttt{aug-cc-pVQZ} (quadruple-zeta) basis.
        \textbf{c}:
        The density total variation contributions along the molecular axis, relative to the GKS (OEP) or CCSD(T) (IKS) reference.
        \textbf{d,e}:
        Same as panels (a,b) for the carbon monoxide (CO) molecule.
        \textbf{f}:
        Same as panel (c) for the carbon monoxide (CO) molecule.
        \textbf{g}:
        The radial dependence and the asymptotic decay of neon (Ne) atom effective potentials.
    }
    \label{fig:small-molecules}
\end{figure*}

Building on Refs.~\cite{yangDirectMethodOptimized2002, wuDirectOptimizationMethod2003}, we view the problem of finding the effective potentials through the lens of nonlinear potential optimization in \cref{eq:oep-variational,eq:iks-data}. Orbitals $\phi[v]$ can be viewed as implicitly parameterized by the potential $v$ itself through \cref{eq:ks-multiplicative}. Solving \cref{eq:ks-multiplicative} with a variational potential $v = v_\theta$ defines the trial orbitals $\phi _\theta$. Parameters $\theta$ are directly optimized. Crucially, our approach does not assume a linear potential expansion using an auxiliary basis. The parameterization $v_\theta$ is completely free. Below, we introduce a physics-informed trial potential that algebraically enforces known asymptotic decay.

Trial potentials must be constrained by known OEP boundary conditions. For neutral molecules and atoms, exact OEPs decay as $v \simeq \nicefrac{-1}{r}$, dominated by the exchange component at long ranges. This decay admits an interpretation of the XC component as a Coulomb potential of an auxiliary charge distribution $n_\xc(\rr)$ through the Poisson equation
\begin{equation}
\label{eq:xc-poisson}
    \laplacian v_\xc (\rr) = -4 \pi \, n_\xc(\rr) \; .
\end{equation}
The asymptotic boundary condition is then equivalent to the normalization condition
\begin{equation}
\label{eq:xc-norm}
    \int \dd[3]{\rr} \; n_\xc (\rr) = -1 \; ,
\end{equation}
setting the first term in the multipole expansion of the total XC potential. Therefore, the total effective charge density $n + n_\xc$ can be seen as the electrostatic source density of the electronic effective potential, including all exchange and correlation effects~\cite{trushinNumericallyStableOptimized2021}, and must integrate to $N-1$. We exploit the bulk-boundary correspondence established by \cref{eq:xc-poisson,eq:xc-norm} to enforce asymptotic conditions.

Some hybrid XC functional approximations use a fraction $\gamma < 1$ of exact exchange. Because exchange effects define the long-range behavior, the asymptotic decay of the XC potential is then modified to $v_\xc \simeq \nicefrac{-\gamma}{r}$, and the corresponding $n_\xc$ from \cref{eq:xc-norm} is required to normalize to $-\gamma$.

For molecular systems studied in this work, we use a real-space trial potential
\begin{equation}
\label{eq:total-ansatz}
    v _\theta (\rr) = v_\ext (\rr) + v_\fa (\rr) + {v} _\ms ^{(\theta)} (\rr) \, ,
\end{equation}
where $v_\ext$ is the fixed background potential and $v_\fa$ is the Fermi-Amaldi (FA)~\cite{wuDirectOptimizationMethod2003} potential generated by a fixed reference density. The final term in \cref{eq:total-ansatz} is the \emph{multipole splat} potential $v_\ms$, with free parameters $\theta$. Because the FA term in \cref{eq:total-ansatz} already carries the correct asymptotic decay, the main role of $v_\ms$ is to correct the bulk.

The multipole splat term $v_\ms$ is represented as a Coulomb potential of a parametrized charge distribution $n_\ms$. The distribution $n_\ms$ is modeled as a superposition of many floating normalized Gaussian monopoles and dipoles with optimizable charges, dipole moments, as well as centers and widths. Each Gaussian charge distribution $\rho _k$ generates a known analytic electrostatic potential $v_k$ by Gauss' theorem, ensuring that $\laplacian v_k = -4 \pi \rho _k$. Optimization details and precise functional forms of $\rho _k$ and $v_k$ are provided in Methods.

The target asymptotic boundary condition $v_\xc \simeq \nicefrac{-\gamma}{r}$ is controlled by the total charge $Q_\ms$ in the multipole splat cloud.
After subtracting the Hartree contribution $v_H$, the decay is fixed by the monopole term in the multipole expansion of the trial XC potential $v_\xc = v_\fa + v_\ms - v_H$, around the arbitrary origin
\begin{equation}
    v_\xc (r) = - \frac{1 - Q_\ms}{r} + \mathcal{O} \left( \frac{1}{r^2} \right) \; ,
\end{equation}
with $r = |\vb{r}|$. This is the reason multipole splats are parameterized to satisfy the exact sum rule constraint
\begin{equation}
\label{eq:sum-rule}
	Q_\ms = 1 - \gamma \; ,
\end{equation}
at all times, with $\gamma$ set to the exact exchange fraction in the given hybrid functional. By building \cref{eq:sum-rule} into the parameterization of the multipole splat cloud, the target asymptotic decay is imposed identically for every parameter value visited during optimization.

\subsection*{Local potential and XC source profiles}

Generalized Kohn-Sham calculations with global hybrid functionals contain a fixed fraction $\gamma$ of exact exchange \cite{perdewRationaleMixingExact1996, kummelOrbitaldependentDensityFunctionals2008a}
\begin{equation}
\label{eq:exx}
    E_\exx = -\frac{1}{2} \sum_{k l} \int \dd[3]{\rr} \int \dd[3]{\rr '} \frac{\phi ^* _k (\rr) \phi ^* _l (\rr ') \phi _l (\rr) \phi _k (\rr ')}{|\rr - \rr'|} \; .
\end{equation}
The EXX functional in \cref{eq:exx} is defined by its direct dependence on occupied orbitals, requiring the OEP approach to restore the Kohn-Sham framework. We study the EXX-only solutions as well as some of the most popular hybrid functionals -- B3LYP~\cite{beckeDensityfunctionalThermochemistryIII1993} and PBE0~\cite{adamoReliableDensityFunctional1999}.

For B3LYP, $\gamma=\nicefrac15$ and for PBE0, $\gamma=\nicefrac14$. Multipole splat potentials decay with the required $\nicefrac{-\gamma}{r}$ because it has been built into the variational potential as a constraint using the sum rule in \cref{eq:sum-rule} rather than treating it as an emergent property. An example of the controllable asymptotic decay can be seen in the right panel of \cref{fig:small-molecules}.

Optimized trial potentials do not differentiate between Hartree, exchange, and correlation contributions, which are commonly separated in DFT literature~\cite{martinElectronicStructureBasic2020} and codes. We isolate the XC component by subtracting the Hartree background at convergence as
\begin{equation}
\label{eq:xc-potential}
	v_\xc (\rr) = v_\fa (\rr) + v_\ms (\rr) - v_H (\rr) \; ,
\end{equation}
where $v_H$ is the Hartree (Coulomb) potential sourced by the converged OEP or IKS density $n$.

Multipole splat potentials recover smooth XC potential spatial profiles. In \cref{fig:small-molecules}, we show the potentials $v_\xc$ and their source densities $n_\xc$, defined by \cref{eq:xc-poisson}, as smooth values along the axis of silicon monoxide (SiO) and carbon monoxide (CO). The converged examples display smooth functions with sharp physical features.

Finite-basis numerical noise amplification is absent in the examples, unlike in some OEP and IKS solvers that rely on auxiliary potential basis sets. The XC source density $n_\xc$ reflects the atomic inter-shell peak structure of the molecule and is a primary output of the multipole splat optimization. Modeling the source directly and integrating it analytically avoids numerical finite-difference laplacians applied to a grid representation of the potential, which commonly introduce a new dominant source of error by amplifying numerical noise.

The accuracy of a given OEP calculation can be evaluated from its GKS counterpart. We focus on three figures of merit: the HOMO eigenvalue $\epsilon_\text{HOMO}$, the localization energy $E_\text{loc} = E_\oep - E_\text{GKS}$, and the density total variation (TV)
\begin{equation}
    \mathcal{D}_\text{TV} = \int \dd[3]{\rr} \; \left| n (\rr) - n_0 (\rr) \right| \; ,
\end{equation}
with respect to the appropriate reference density $n_0$ taken from the GKS calculation. For the IKS case, the reference density $n_0$ is taken directly from an independent CCSD(T) calculation.

\begin{figure*}[t]
    \centering
    \includegraphics[width=\linewidth]{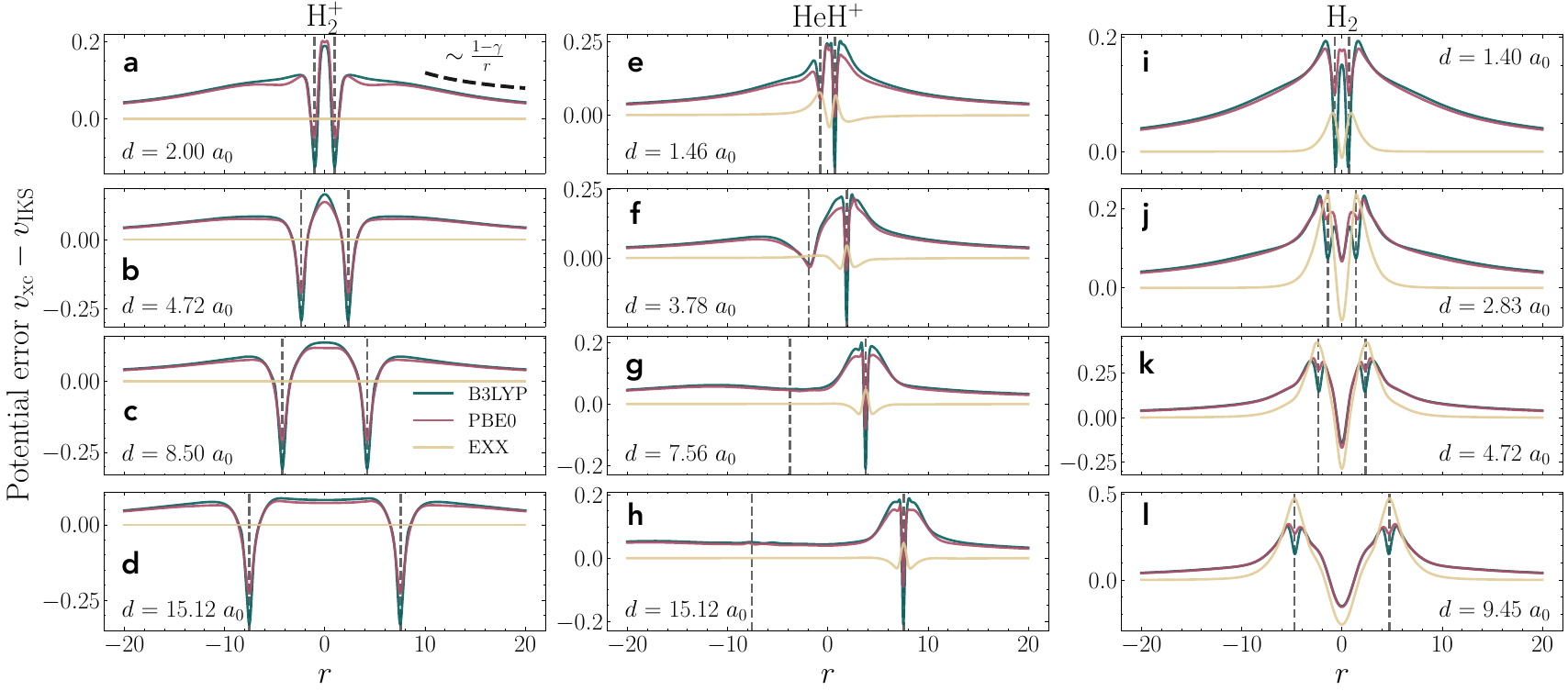}
    \caption{
        \textbf{Spatially resolved exchange-correlation potential errors.}
        All quantities given in Hartree atomic units.
        \textbf{a-d}:
        Differences between B3LYP, PBE0 and EXX OEPs and the IKS reference potentials, defined in \cref{eq:potential-err}, for the dissociating $\text{H}_2^+$ molecule. Only the XC components are compared. The reference is obtained by inverting a CCSD(T) density. Profiles are plotted along the molecular axis $r$. Internuclear separation $d$ increases downward and is labeled in each panel. Orbitals are represented in the \texttt{aug-cc-pV5Z} basis. Vertical dashed lines mark the nuclei (He is on the left in $\text{HeH}^+$). The black dashed curve indicates the $\nicefrac{(1-\gamma)}{r}$ asymptotic form, where $\gamma$ is the exact-exchange fraction. Vertical scales vary between panels.
        \textbf{e-h}:
        Same as panels a-d for the $\text{HeH}^+$ molecule.
        \textbf{i-l}:
        Same as panels a-d for the $\text{H}_2$ molecule.
    }
    \label{fig:dissociation}
\end{figure*}

The OEP problem can be seen as the GKS optimization under the constraint of having local potentials. Therefore, $E_\oep \geq E_\text{GKS}$ holds in all cases. The difference $E_\text{loc} \geq 0$ is the energy penalty of localizing the potential. However, numerical values should be compared with caution for two distinct reasons. Firstly, the $E_\text{GKS}$ floor is a non-monotonically moving target for growing finite orbital basis sets. Secondly, mismatched expressivity between finite parameterizations of orbitals and the potential invalidates the Hohenberg-Kohn one-to-one correspondence between potentials and densities. This mismatch can lead to unphysical potentials that minimize the energy at the expense of representation blind spots instead of building real physical features \cite{staroverovOptimizedEffectivePotentials2006, gorlingRelationExchangeonlyOptimized2008, trushinNumericallyStableOptimized2021}. In practice, low localization energy is a necessary but not sufficient condition for success \cite{trushinNumericallyStableOptimized2021} and has to be supplemented by independent visual and numerical checks.

The OEP density TV should be treated similarly. It should attain low per-electron values but some differences are expected with respect to GKS because energy is being minimized, not density divergence. The IKS case is simpler -- lower TV values indicate a closer density match, given a reliable reference density.

Multipole splat potentials achieve milli-Hartree localization energies while maintaining a close match with GKS densities, as measured by the TV. Exact numerical values for both figures of merit can be found in \sref{tab:energy-errs}. Physically meaningful potential, source density and spatial TV profiles on \cref{fig:small-molecules} demonstrate the absence of numerical problems. Additional potential profiles are shown in \sref{fig:potential-gallery}.

\subsection*{Spatially resolved exchange-correlation potential errors}

Local OEPs of orbital-dependent XC approximations can be directly compared to precise IKS references. The multipole splat solver provides both ingredients. In this section, we compare hybrid potentials against the IKS potential obtained from a CCSD(T) density. We define the spatially resolved XC potential error as
\begin{equation}
\label{eq:potential-err}
    \Delta v_\xc^F (\rr) = v_\xc^F (\rr) - v_\xc^\iks (\rr) \; ,
\end{equation}
where $F \in \{ \mathrm{EXX},\, \mathrm{B3LYP},\, \mathrm{PBE0} \}$ denotes the XC functional approximation. All potentials vanish at infinity, fixing their additive constants. We independently subtract each calculation's own Hartree potential to extract its XC component, as per \cref{eq:xc-potential}. The comparison includes density relaxation and does not isolate a pure correlation error. The IKS result is treated in a finite multipole splat representation, not an exact continuum potential.

\cref{fig:dissociation} follows these errors during dissociation of $\text{H}_2^+$, $\text{HeH}^+$ and $\text{H}_2$. The three systems probe one-electron cancellation, asymmetric dissociation and bond breaking. The asymptotic part of the comparison is fixed by construction: subtracting the $\nicefrac{-1}{r}$ IKS tail from a hybrid $\nicefrac{-\gamma}{r}$ tail gives $\Delta v_\xc^F \simeq \nicefrac{(1-\gamma)}{r}$. EXX has no such leading tail error. This constraint does not determine the sign or magnitude of errors within the molecule.

Correlation has to vanish identically in $\text{H}_2^+$ because it is a single-electron system. Exact exchange cancels the Hartree self-interaction~\cite{perdewSelfinteractionCorrectionDensityfunctional1981} while B3LYP and PBE0 cannot. The EXX error in the left column of \cref{fig:dissociation} is indistinguishable from zero on the plotted scale at every separation. B3LYP and PBE0 show similar error profiles throughout dissociation. Both have negative minima near the nuclei and positive errors in the bond and outer regions. The nuclear minima separate as the bond stretches, leaving a broad positive plateau between them. B3LYP has deeper nuclear minima than PBE0, but the spatial pattern is shared. The one-electron comparison exposes incomplete self-interaction cancellation within the molecule as well as in the asymptotic tail.

The middle column follows $\text{HeH}^+$ toward dissociation into neutral He and a bare proton. This limit tests the tendency of approximate functionals to favor fractional fragment charges~\cite{perdewDensityFunctionalTheoryFractional1982, cohenInsightsCurrentLimitations2008}. At large separations, the strongest hybrid potential errors remain near the proton: a broad positive feature surrounds a narrow negative minimum. EXX stays closer to the reference in this region. The residual identifies where the potentials disagree, but does not by itself establish a fractional charge or a missing interfragment step. The density near the proton becomes small during dissociation, making the inverted potential there less well constrained.

\begin{figure*}[!t]
    \centering
    \includegraphics[width=\linewidth]{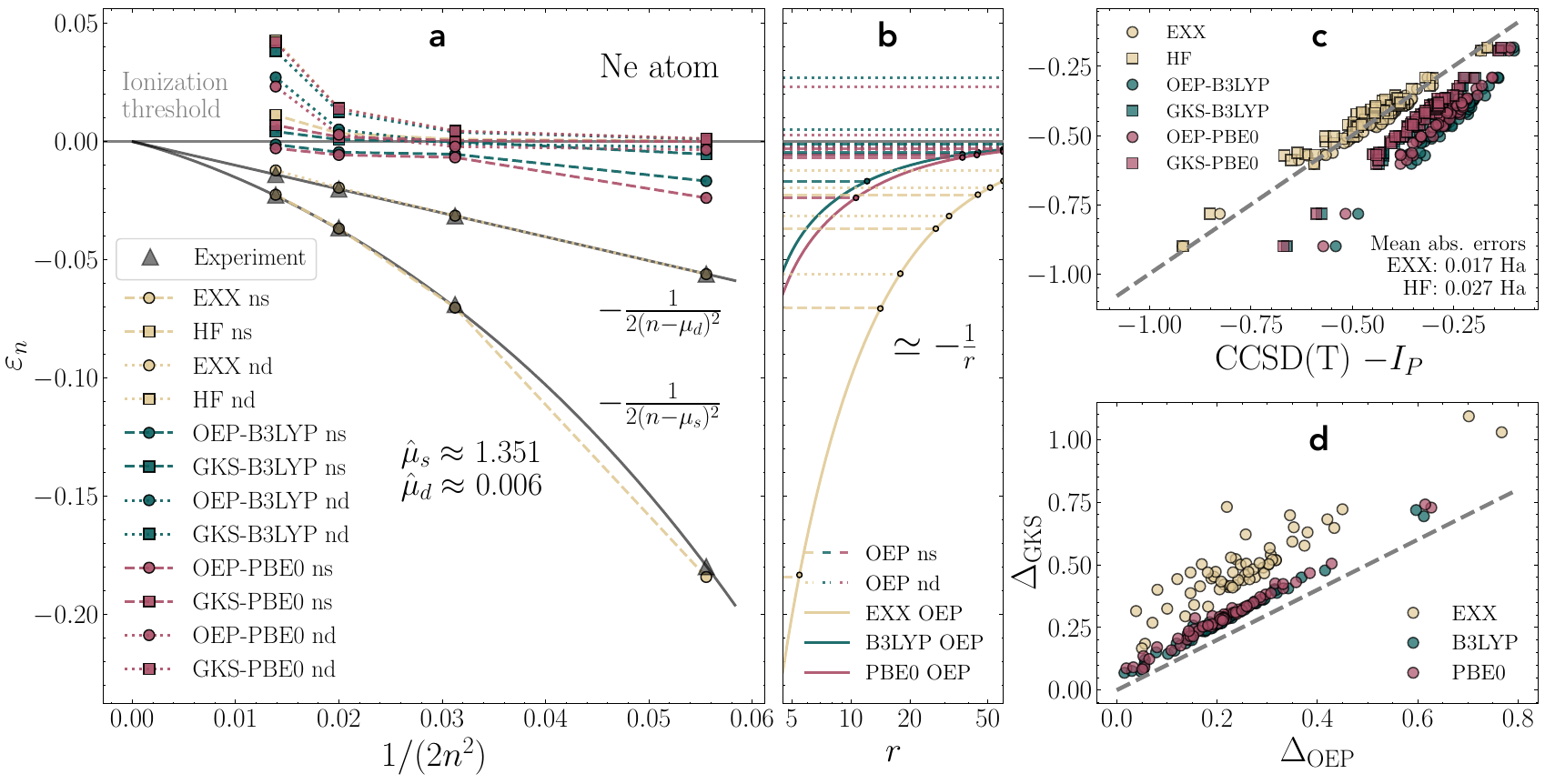}
    \caption{
        \textbf{Atomic excited states and gaps.}
        All quantities given in Hartree atomic units.
        \textbf{a}: Neon atom Rydberg $ns$ ($l=0$) and $nd$ ($l=2$) series $\varepsilon_n$, plotted against $\nicefrac{1}{2n^2}$. Experimental reference energies relative to the ionization limit and EXX-OEP quantum-defect fits $\Hat \mu _s \approx 1.351$ and $\Hat \mu _d \approx 0.006$ from \cref{eq:quantum-defect} are highlighted. OEP eigenvalues and their GKS or Hartree-Fock (HF) counterparts are plotted as well.
        \textbf{b}: Converged OEP potentials on a logarithmic radial scale, sharing the energy axis with panel (a). Horizontal lines mark the $ns$ and $nd$ eigenvalues, with classical turning points marked.
        \textbf{c}: HOMO eigenvalues against negative CCSD(T) ionization potentials $-I_P$ for the selected 60 GW100 systems with $N \leq 40$ electrons. Mean absolute errors $|\varepsilon _\text{HOMO} + I_P |$ for EXX ($\approx 17$ mHa) and HF ($\approx 27$ mHa) are shown, averaged over these systems.
        \textbf{d}: OEP vs. GKS HOMO-LUMO gaps $\Delta$ for the same set.
    }
    \label{fig:spectrum}
\end{figure*}

The right column of \cref{fig:dissociation} shows a different pattern as $\text{H}_2$ enters the strongly correlated bond-breaking regime. At $1.50 \, \mathring{\text{A}}$, the hybrid midpoint errors remain positive while the EXX error is negative. At $2.50$ and $5.00 \, \mathring{\text{A}}$, all three approximations underestimate the reference XC potential at the midpoint and overestimate it in the flanking regions near the nuclei. The midpoint structure is therefore too low relative to its surroundings. B3LYP and PBE0 reduce both the midpoint deficit and the flanking excesses relative to EXX at these larger separations, but do not remove them. In these hybrids, reduced bond-region errors coexist with an asymptotic error set by the reduced exact-exchange fraction.

The three dissociations separate asymptotic accuracy from accuracy within the molecular region. EXX agrees with the one-electron reference but develops substantial bond-region errors in stretched $\text{H}_2$ despite its correct tail. Hybrid corrections reduce some of those errors but introduce others, most clearly in $\text{H}_2^+$. The local-potential comparison resolves \emph{where} any given XC approximation improves agreement with an accurate correlated reference as well as the spatial distribution of errors.

\subsection*{Spectral properties and Rydberg physics}

In the Kohn-Sham framework, occupied and unoccupied orbitals are eigenstates of the same local potential. In GKS or Hartree-Fock (HF) theory, the nonlocal exchange operator acts differently on the two subspaces. An electron in a virtual HF orbital would experience the field of all $N$ electrons instead of $N-1$. The residual attraction it feels decays exponentially instead of as $\nicefrac{-1}{r}$. Therefore, the unoccupied spectrum is a probe of local potentials.

The long Coulomb-like tail of the exact KS potential of a neutral atom supports an infinite Rydberg series of orbital eigenvalues converging to zero, the continuum threshold. Their energies follow the form
\begin{equation}
\label{eq:quantum-defect}
    \varepsilon_n \approx -\frac{1}{2(n-\mu_l)^2} \; ,
\end{equation}
where $n$ is the principal quantum number and $\mu_l$ is the \emph{quantum-defect}~\cite{seatonQuantumDefectTheory1983} associated with angular momentum $l$. The quantum defect depends on how far a Rydberg electron penetrates the atomic core -- it is large for low-$l$ states that penetrate the core and approximately vanishes for high-$l$ states held out by the centrifugal barrier, which see a bare $\nicefrac{-1}{r}$ tail and remain hydrogenic.

The left panel of \cref{fig:spectrum} shows the $ns$ ($l=0$) and $nd$ ($l=2$) Rydberg series of the Ne atom, plotted against $\nicefrac{1}{(2n^2)}$ so that a purely hydrogenic series falls on a straight line through the origin. We compare the orbital eigenvalues with experimental excitation energies measured relative to the ionization limit, $E_n^\text{NIST} - I_P$~\cite{kramidaNISTAtomicSpectra1999}, as detailed in \sref{sec:details}. EXX-OEP closely follows the NIST reference levels~\cite{kramidaNISTAtomicSpectra1999} across both series and principal quantum numbers shown here, with no empirical asymptotic corrections. These unoccupied KS eigenvalues are not many-body excitation energies, but their quantum defects do provide a diagnostic of the local potential. Despite treating exchange exactly, HF does not reproduce the bound series because its virtual orbitals do not see the $\nicefrac{-1}{r}$ tail. This comparison demonstrates that the local potential and the correct asymptotics are both necessary ingredients. The hybrid OEP levels are substantially less bound than the reference levels, consistent with their weaker $\nicefrac{-\gamma}{r}$ tails. In the complete-basis limit, these tails still support infinitely many bound states~\cite{seatonQuantumDefectTheory1983}. Our results show that some computed levels lie above the continuum threshold when using finite bases.

On the center panel of \cref{fig:spectrum}, converged OEP potentials are drawn against radial distance, with each $ns$ and $nd$ eigenvalue overlaid as a horizontal line and its classical turning point marked. Each spectral line in the left panel maps directly onto the depth and width of the supporting potential well. The EXX-OEP potential tracks $\nicefrac{-1}{r}$ out to $50$ Bohr, while the hybrid potentials remain shallower and approach their imposed $\nicefrac{-\gamma}{r}$ tails.

Resolving Rydberg physics requires diffuse orbital basis functions. We augment the standard basis with additional diffuse Gaussians for the Ne spectrum (see \sref{tab:rydberg-basis}). Multipole splats accommodate the same physics without modification. Parameterized exponents in \sref{eq:exponent-params} allow splats to describe spatial structure over many length scales while retaining sharp core features. The asymptotic tail is fixed independently by the charge constraint in \cref{eq:sum-rule}. Fitting \cref{eq:quantum-defect} to the converged EXX-OEP levels yields $\mu_s = 1.351$ and $\mu_d = 0.006$, against $1.331$ and $0.012$ extracted from Ref.~\cite{kramidaNISTAtomicSpectra1999}. Both series are recovered to within $5$ mHa at $n=3$ by multipole splats from first principles.

Occupied-state spectra are constrained as well. For the exact KS system, the ionization-potential theorem gives $\epsilon_\text{HOMO} = -I_P$~\cite{perdewDensityFunctionalTheoryFractional1982}, for the molecular ionization potential $I_P$. This equality is not exact for the approximate EXX and hybrid functionals studied here, making the ionization potential an independent diagnostic. The upper right panel of \cref{fig:spectrum} compares HOMO eigenvalues against CCSD(T) ionization potentials across 60 molecules in the GW100~\cite{vansettenGW100BenchmarkingG0W02015} dataset, with at most 40 electrons. Orbitals were represented using the \texttt{aug-cc-pVTZ} orbital basis. EXX-OEP tracks the reference closely over the full $\sim 1$ Ha range spanned by the dataset, while the hybrids deviate systematically, by an amount that grows as the exact-exchange fraction $\gamma$ gets smaller.

Independent numerical benchmarks against real-space EXX-OEP reference eigenvalues \cite{trushinNumericallyStableOptimized2021} are reported in \sref{tab:method-comparison}, with methodological details and potential profiles in \sref{sec:comparison} and \sref{fig:comparison}. Across the five atoms with available references, the mean absolute deviation of the multipole-splat HOMO eigenvalues is $7.1$ mHa, without enforcing a HOMO condition or introducing any spectral target into the optimization.

The lower right panel of \cref{fig:spectrum} compares OEP and GKS gaps. The OEP gap $\Delta_\oep$ is smaller than the GKS gap $\Delta_\text{GKS}$ throughout the dataset. This difference reflects the spectral rearrangement caused by replacing the nonlocal GKS operator with a local multiplicative potential in \cref{eq:ks-multiplicative}. Among the approximations studied, EXX receives the largest correction. In exact KS theory, the fundamental gap differs from the orbital gap by the XC derivative discontinuity~\cite{perdewDensityFunctionalTheoryFractional1982, shamDensityFunctionalTheoryEnergy1983, perdewPhysicalContentExact1983}. GKS eigenvalue gaps can incorporate part of this contribution~\cite{seidlGeneralizedKohnShamSchemes1996, perdewUnderstandingBandGaps2017}. This provides context for the gap comparison, but the present gap difference alone is not sufficient to estimate the derivative discontinuity. Regardless, these potentials may offer a principled starting point for linear-response TD-DFT, where excitation energies also depend on the XC kernel.

\subsection*{Larger molecular systems}

\begin{figure}[t]
    \centering
    \includegraphics[width=\linewidth]{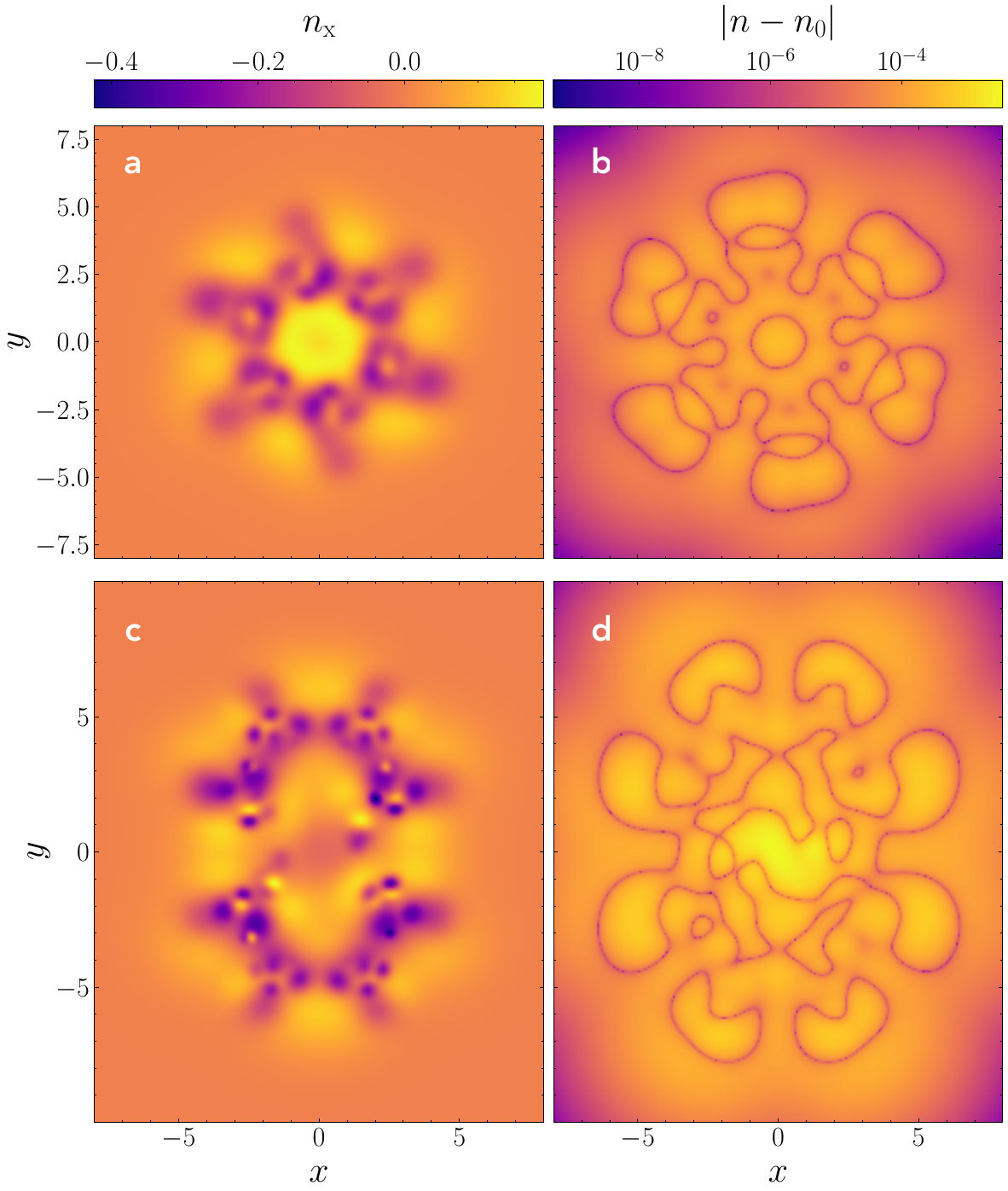}
    \caption{
        \textbf{Exact exchange OEP source densities and density errors for benzene and naphthalene.}
        All quantities given in Hartree atomic units.
        \textbf{a,c}: The exchange source density for benzene (a) and naphthalene (c), evaluated on a plane $1$ Bohr above the molecular plane.
        \textbf{b,d}: The pointwise density variation $\left| n - n_0 \right|$ against the GKS reference $n_0$ for benzene (b) and naphthalene (d).
    }
    \label{fig:large}
\end{figure}

Effective potentials are usually expanded in an auxiliary basis whose balance against the orbital basis has to be maintained by hand~\cite{staroverovOptimizedEffectivePotentials2006, heaton-burgessOptimizedEffectivePotentials2007, trushinNumericallyStableOptimized2021}. Splats are allocated per occupied orbital (see \sref{sec:params}), so the parameterization needs no per-system tuning.

In this section, we compute EXX OEP potentials for benzene (42 electrons) and naphthalene (68 electrons), shown in \cref{fig:large} on a plane $1$ Bohr above the molecular plane, using the \texttt{cc-pVTZ} orbital basis. The potentials resolve $\pi$-conjugated systems. For benzene, exchange charge is distributed evenly around the ring, the real-space signature of complete $\pi$ delocalization. Splats are initialized symmetrically with respect to the molecular point group and optimized without symmetry constraints. The converged exchange potential source density preserves the six-fold symmetry. On the other hand, naphthalene breaks the pattern. The map separates the peripheral carbons from the ring-fusion region, consistent with the bond inequivalence that distinguishes naphthalene from benzene and absent from the smaller molecule.

Accuracy does not degrade with system size for these examples, provided that enough splats are used. The right column of \cref{fig:large} shows the pointwise density deviation from the GKS reference staying below $0.005$ a.u. across both molecules, with nodal lines where the two densities cross appearing as sharp minima. Similar to a standard GKS calculation, the per-step cost of direct potential optimization is dominated by the integral evaluation and eigendecomposition steps.

\section*{Discussion}

Effective single-electron potentials are naturally defined as solutions to constrained optimization problems. Their treatment as implicit density functionals within the self-consistent field (SCF) framework~\cite{sharpVariationalApproachUnipotential1953, talmanOptimizedEffectiveAtomic1976, kummelOptimizedEffectivePotential2003} commonly requires an ill-conditioned inversion of the density response operator whose regularization has occupied the field for decades~\cite{staroverovOptimizedEffectivePotentials2006, heaton-burgessOptimizedEffectivePotentials2007, trushinNumericallyStableOptimized2021}, blocking access to force fields and spectra furnished by multiplicative potentials. In Methods, we formulate OEP and IKS problems as nonlinear optimization~\cite{yangDirectMethodOptimized2002, wuDirectOptimizationMethod2003} over trial potentials, replacing response inversion with first-order optimization. We explore this tradeoff and argue that modern nonlinear optimization has become a practical tool for reliable computation of effective potentials.

Multipole splats have been proposed as a natural and expressive choice for the nonlinear trial potential parameterization, with exact asymptotics built in and guaranteed. They have been inspired by modern methods for three-dimensional scene reconstruction in computer vision~\cite{kerbl3DGaussianSplatting2023}. A robust first-order gradient-based optimizer has been borrowed from the AI literature. These ingredients are ready to be used as self-contained OEP and IKS solvers, removing the need to hand-tune auxiliary bases and asymptotic corrections, and giving functional designers more spatial and geometrical diagnostics.

To complete the IKS theory frame, we have shown that KS potential inversion through WY~\cite{wuDirectOptimizationMethod2003} maximization is equivalent to maximum-likelihood estimation~\cite{jaynesInformationTheoryStatistical1957} of the full equilibrium KS density operator using the density as the sufficient statistic~\cite{hasegawaExponentialMixtureFamilies1997}. This theoretical connection roots the well-known mathematical properties and stability of the WY functional~\cite{liebDensityFunctionalsCoulomb1983} in statistical learning theory \cite{wainwrightGraphicalModelsExponential2008} and allows us to extend the WY inversion protocol to finite-temperature states. These connections position IKS as a tool for building the next generation of electronic structure datasets.

Comparing self-consistent XC potentials against IKS references from the same solver offers real-space resolution of approximate functional errors~\cite{cohenInsightsCurrentLimitations2008}. The dissociation profiles show that correct asymptotics do not guarantee accuracy within the molecule. Hybrids reduce some bond-region errors in stretched $\text{H}_2$ despite their incorrect tails. The exact exchange potential closely follows reference Rydberg series and quantum defects~\cite{kramidaNISTAtomicSpectra1999}, showing the role of both locality and asymptotic decay in the unoccupied KS spectrum. The same optimization produces accurate HOMO eigenvalues across the GW100 subset~\cite{vansettenGW100BenchmarkingG0W02015} and exposes the gap redistribution associated with replacing a nonlocal operator by a local potential~\cite{perdewDensityFunctionalTheoryFractional1982, seidlGeneralizedKohnShamSchemes1996}, while exchange source densities for $\pi$-conjugated molecules demonstrate robust results at the scale of tens of electrons.

Some numerical scaffolding is still required. Non-uniqueness caused by finite orbital and potential representations is the main obstacle to completely unregularized determination of OEPs. This is a well-known consequence of Hohenberg-Kohn~\cite{hohenbergInhomogeneousElectronGas1964} theorems breaking in finite bases~\cite{staroverovOptimizedEffectivePotentials2006, gorlingRelationExchangeonlyOptimized2008}. In practice, regularization is effective in mitigating this problem, but it cannot be removed completely. Enforcing HOMO energy conditions can be added as extra regularization~\cite{trushinNumericallyStableOptimized2021}. Additionally, first-order optimization takes many iterations to converge and analytic three-center integral evaluation remains CPU-bound. Immediate next steps are natural gradient optimization~\cite{amariNaturalGradientWorks1998} and a GPU-resident integral engine, which would remove host-device communication overhead. We expect these extensions to make the execution speed competitive with SCF methods.

Having established a close connection between the KS unoccupied states, the self-interaction error, and potential asymptotics, we plan on using multipole splats with exact asymptotics on functionals that otherwise yield incorrectly decaying potentials, building algebraically constrained adaptive range separation~\cite{liuImprovingPerformanceLongRangeCorrected2017}. Enforcing long-range behavior via basis preprocessing has already been shown to meaningfully correct spectral properties in Ref.~\cite{trushinImprovingExchangeCorrelationPotentials2025}. The learning setup is agnostic to the theory supplying the energy or the density, opening a route to orbital-dependent correlation functionals whose self-consistent potentials are only now becoming available~\cite{trushinAccurateCorrelationPotentials2025}. When a stationary total-energy functional derived from a self-energy approximation is available, its local OEP can be computed as well. An extension of multipole splats to periodic systems is possible in reciprocal space as well \cite{elsborgGlobalPlaneWaves2026}. In addition, direct amortization of the proposed solvers can be achieved by predicting splat parameters directly~\cite{elsborgELECTRACartesianNetwork2026} using modern neural networks. Amortization may reduce the need to run independent calculations during TD-DFT evolution by effectively storing interpolated results in neural network parameters \cite{wu4DGaussianSplatting2024}. A model predicting splat parameters inherits the asymptotic decay by construction.

Inverse Kohn-Sham potentials can also be retrieved from state-of-the-art synthetic and experimental data using the proposed generalization of the finite-temperature~\cite{merminThermalPropertiesInhomogeneous1965a} WY objective. The exact convex loss gradient only needs point-wise density residuals. Therefore, different approaches like NQS or cold atom experiments~\cite{mazurenkoColdatomFermiHubbard2017, khatamiVisualizingStrangeMetallic2020}, where sampling is much easier than direct evaluation, are accommodated with no reformulation. This framework steps away from XC functional approximations where data is available.

In practice, the numerical reliability of multipole splat optimization is a prerequisite both for studies of exchange-correlation physics and for the AI training that functional development requires. We expect the framework and its natural extensions to be productive on both sides of this correspondence.

\section*{Methods}

\subsection*{Local and nonlocal potentials}

The Kohn-Sham potential $v$ in \cref{eq:ks-multiplicative} is traditionally decomposed as $v = v_\ext + v_H + v_\xc$, where $v_\ext$ is the external potential, $v_H$ is the classical Coulomb (Hartree) contribution, and $v_\xc$ is the exchange-correlation component. The correlated character of the final solution is built in through the density dependence of the XC potential, modeled by the XC energy functional $E_\xc [n]$ as $v_\xc = \fdv{E_\xc}{n}$. Densities calculated from solutions to \cref{eq:ks-multiplicative} are called $v$-representable because they are compatible with a multiplicative scalar potential $v$.

The exact XC potential comes with theoretical guarantees. The exact highest occupied molecular orbital KS eigenvalue is identified with the negative ionization potential of the molecule~\cite{perdewDensityFunctionalTheoryFractional1982}, $\epsilon _\text{HOMO} = -I_P$. The XC potential decays as $v_\xc \simeq -\nicefrac{1}{r}$ asymptotically as $r \rightarrow \infty$ and is free from the self-interaction error. SIE cancelation is crucial for localized states, transition metal oxides, as well as time-dependent and response properties in general.

Approximate XC energy functionals and finite basis expansions~\cite{linMathematicalIntroductionElectronic2019} spoil these exact guarantees in practical calculations. When approximate XC energy functionals are modeled as true density functionals, $E_\xc = E_\xc [n]$, potentials can be obtained by direct functional differentiation. However, modern hybrid functionals explicitly depend on orbitals, $E_\xc = E_\xc [\phi]$.

The resulting stationary-point generalized KS equations~\cite{kummelOrbitaldependentDensityFunctionals2008a}, $\fdv{E}{\phi} = 0$, include non-local contributions to the potential. For example, hybrid functionals contain a fraction $\gamma$ of the nonlocal exact exchange energy, $E_\xc = E _\text{loc} [n] + \gamma E_\exx [\phi]$. The GKS orbitals then solve
\begin{equation}
\label{eq:gks-hybrid}
    \left(-\tfrac12 \laplacian + v _\text{loc} (\rr) \right) \phi _k (\rr) - \gamma (K \phi) _k (\rr) = \epsilon _k \phi _k (\rr) \, ,
\end{equation}
where $v_\text{loc}$ is the local component of the potential and $K$ is the nonlocal (orbital-mixing) exchange potential
\begin{equation}
	(K \phi) _k (\rr) = \sum _l \int \dd[3]{\rr '} \frac{\phi _k (\rr') \phi _l (\rr')}{|\rr - \rr '|} \phi _l (\rr)
\end{equation}
formed from occupied KS orbitals. Mathematically, \cref{eq:gks-hybrid} simply admits more solutions than \cref{eq:ks-multiplicative}, departing from the $v$-representable context~\cite{kummelOptimizedEffectivePotential2003}.

If we attempt to compute the KS potential by direct differentiation of an orbital-dependent $E_\xc$, we reach a roadblock. Schematically,
\begin{equation}
    v_\xc = \fdv{E_\xc [\phi]}{n} = \fdv{E_\xc [\phi]}{\phi} \cdot \fdv{\phi}{n} \; .
\end{equation}
Because the map from occupied orbitals to the density is many-to-one, including the freedom to rotate the occupied orbitals, $\fdv{\phi}{n}$ is not uniquely defined without specifying an inverse density-to-potential map and an orbital gauge. In principle, $v_\xc$ can be obtained as a solution of a linear system~\cite{sharpVariationalApproachUnipotential1953, talmanOptimizedEffectiveAtomic1976, kummelOptimizedEffectivePotential2003}
\begin{equation}
\label{eq:oep-linear}
    \int \dd[3]{\rr '} \chi _s (\rr, \rr ') v_\xc (\rr') = t(\rr)
\end{equation}
where $\chi _s = \fdv{n}{v}$ is the KS static response function and $t=\fdv{E_\xc}{v}$. However, inversion of \cref{eq:oep-linear} is notoriously ill-conditioned in finite basis sets~\cite{trushinNumericallyStableOptimized2021, trushinAvoidingSpinContamination2023}, costing accuracy at best and causing uncontrolled numerical instabilities at worst.

\subsection*{Unified framework for optimized and inverted potentials}

The KS orbitals remain expressed as a linear combination of suitably chosen basis functions $\chi _\mu$ indexed by Greek indices $\mu$
\begin{equation}
\label{eq:variational-orbitals}
    \phi _k (\rr) = \sum _\mu C _{\mu k} [v] \, \chi _\mu (\rr) \; ,
\end{equation}
by viewing the linear combination coefficients $C$ as a functional of the trial potential $v$ or, equivalently, as a function of any variational parameters $\theta$. Coefficients $C$ are determined as a solution to \cref{eq:ks-multiplicative} in the chosen basis set $\chi _\mu$, giving rise to the standard generalized eigenvalue problem of the form $H C = S C \epsilon$. For non-degenerate eigenvalues, the eigendecomposition in a finite basis is a differentiable mapping $H \mapsto (C, \epsilon)$ with gradients given by well-known first-order perturbation theory formulas \cite{sakuraiModernQuantumMechanics2017}. For degenerate eigenvalues, we establish a heuristic regularization procedure outlined in \sref{sec:gradients}, implemented as a custom gradient rule in our differentiable solver.

The effective single-particle Hamiltonian $H$ is simply the operator
\begin{equation}
\label{eq:effective-hamiltonian}
    H _\theta = -\tfrac{1}{2} \laplacian + v _\theta \; ,
\end{equation}
projected down into the orbital basis $\chi$. The variational potential $v_\theta$ contributes to the eigenvalue equation as an additive term with its matrix representation
\begin{equation}
\label{eq:potential-basis}
    V _{\mu \nu} = \int \dd[3]{\rr} \; v_\theta (\rr) \, \chi _\mu (\rr) \, \chi _\nu (\rr) \; .
\end{equation}
The projection in \cref{eq:potential-basis} can be viewed as generalized \textit{splatting} onto a finite-dimensional basis set. Similar to the Gaussian splatting method in computer vision literature \cite{kerbl3DGaussianSplatting2023}, multipole splats are optimized exclusively through their projections onto the orbital basis $\chi$. Despite being optimized through generalized splatting defined in \cref{eq:potential-basis}, the potential inherits the real-space asymptotic constraints from our multipole splat ansatz introduced below.

For the OEP, we seek to minimize the GKS energy $E_\text{GKS}$ such that $\phi [v]$ are the $v$-representable orbitals implicitly defined by the multiplicative local potential $v$ in \cref{eq:ks-multiplicative}. In practice, we optimize the regularized variational energy functional
\begin{equation}
\label{eq:total-energy-reg}
    E(\theta) = E _\text{GKS} [\phi[v_\theta]] + \lambda \, R[v_\theta] \;,
\end{equation}
to compute the OEP or the regularized divergence with respect to the reference density $n_0$
\begin{equation}
\label{eq:total-kl-reg}
    \mathcal{D}(\theta) = \mathcal{D} [n [v_\theta], n_0] + \lambda \, R[v_\theta] \;,
\end{equation}
with a small positive constant $\lambda$. The regularization term $R[v]$ in \cref{eq:total-energy-reg,eq:total-kl-reg} is chosen to resolve the null-space of different solutions in favor of smoother potentials. Our specific choice of $R[v]$ is introduced below.

The mapping $\theta \mapsto E(\theta)$ is differentiable because it is built as a composition of differentiable functions. Therefore, optimal variational parameters $\theta$ are computed using gradient-based optimization of the total energy functional in \cref{eq:total-energy-reg}. Parameter gradients require the evaluation of the orbital response function because
\begin{equation}
	\fdv{E}{v_\theta} = \fdv{E[\phi]}{\phi} \cdot \fdv{\phi}{v_\theta} \; .
\end{equation}
The orbital response $\fdv{\phi}{v}$ can be obtained through first-order perturbation theory. Crucially, inverting the response function is not required, removing the main source of numerical instabilities in traditional approaches~\cite{kummelOrbitaldependentDensityFunctionals2008a}. The internal sum over all unoccupied states can be carried out directly or implicitly through the Sternheimer equation~\cite{andradeTimedependentDensityFunctional2007}.

\subsection*{Inverse Kohn-Sham as supervised Hamiltonian learning}

To compute the IKS potential, we start from the quantum Kullback-Leibler (KL) divergence or the relative von Neumann entropy. For a reference state $\Hat \rho _0$ with density $n_0$ and the finite-temperature KS Gibbs state $\Hat \rho _\beta [v]$, the quantum KL divergence is defined as
\begin{equation}
\label{eq:qkl}
    \mathcal{D} _\text{KL} (\Hat \rho_0 \parallel \Hat \rho _\beta[v] ) = - \Tr{ \Hat \rho_0 \left( \ln \Hat \rho _\beta[v] - \ln \Hat \rho _0 \right) }
\end{equation}
where the two states are taken to have the same expected particle number. Only the reference density $n_0$ enters the potential-dependent part of \cref{eq:qkl}, so the full reference state does not need to be constructed. In this work, $n_0$ is obtained using the relaxed CCSD(T)~\cite{bartlettCoupledclusterTheoryQuantum2007}, but reference densities could be imported from arbitrary sources.

We show that minimizing the scaled zero-temperature limit of \cref{eq:qkl} is equivalent to maximizing the WY functional introduced in Ref.~\cite{wuDirectOptimizationMethod2003},
\begin{equation}
\label{eq:wy-functional}
    W[v] = T_s[\phi[v]] + \int \dd[3]{\rr} v (\rr) ( n[v] (\rr) - n_0 (\rr)) \; ,
\end{equation}
where $T_s[\phi]$ is the non-interacting kinetic energy. \cref{eq:qkl,eq:wy-functional} frame a fundamental result in the IKS theory as a supervised Hamiltonian learning problem. Details can be found in \sref{sec:gradients} where we show that
\begin{equation}
\label{eq:kl-equivalence}
    \argmin _v \lim _{\beta \rightarrow \infty} \beta ^{-1} \mathcal{D} _\text{KL} (\Hat \rho _0 \parallel \Hat \rho _\beta [v])  = \argmax _v W[v] \; .
\end{equation}
The result in \cref{eq:kl-equivalence} can be used to generalize the WY IKS approach to finite temperatures. In practice, we minimize the regularized loss
\begin{equation}
\label{eq:total-wy-reg}
    \mathcal{D}(\theta) = -W[v_\theta] + \lambda \, R[v_\theta] \; ,
\end{equation}
inheriting mathematical properties from the original formulation in \cref{eq:qkl}.

The thermal density operator $\Hat \rho _\beta[v] \propto e^{-\beta ( \Hat H[v] - \mu \Hat N )}$ belongs to the \textit{quantum exponential family}~\cite{wainwrightGraphicalModelsExponential2008, hasegawaExponentialMixtureFamilies1997} for a KS Hamiltonian in \cref{eq:effective-hamiltonian}. The potential $v$ is a natural parameter and the density $n$ is the mean parameter (corresponding to a sufficient statistic) of $\Hat \rho _\beta$ so that minimizing \cref{eq:qkl} is maximum-likelihood estimation~\cite{jaynesInformationTheoryStatistical1957, anshuSampleefficientLearningInteracting2021} in that family. Convexity of the log-partition function in the natural parameter is the origin of the convexity of the loss and of the negative semi-definiteness of the KS response $\chi _s$, while its derivative yields the density $n$. At zero temperature this is precisely the Legendre structure identified by Lieb~\cite{liebDensityFunctionalsCoulomb1983}.

Therefore, the WY functional is concave in the variational potential, while the negative WY loss used in \cref{eq:total-wy-reg} is convex in the potential. From \cref{eq:wy-functional}, gradients of the WY functional simply evaluate to the density residual
\begin{equation}
    \fdv{W}{v} = n[v] - n_0 \; ,
\end{equation}
vanishing exactly when $n[v] = n_0$. Therefore, exploiting $n_0$ allows us to bypass the calculation of the unoccupied states appearing in gradients of \cref{eq:total-energy-reg}.

\subsection*{Practical optimization of multipole splats}

We use a real-space trial potential introduced in \cref{eq:total-ansatz} for an isolated molecule with $A$ atoms with charges $Z_a$ at coordinates $\RR _a$ ($a=1, \, \ldots ,\, A$) and $N$ electrons. The fixed background nuclear Coulomb contribution is $v_\ext (\rr) = - \sum _a \flatfrac{Z_a}{|\rr - \RR_a|}$ and
\begin{equation}
    v_\fa (\rr) = \frac{N-1}{N} \int \dd[3]{\rr'} \, \frac{n _\text{ref} (\rr ')}{|\rr - \rr '|}
\end{equation}
is the Fermi-Amaldi potential generated by a fixed reference density $n _\text{ref}$. The final term in \cref{eq:total-ansatz} is the multipole splat potential $v_\ms$, detailed below.

The multipole splat source density $n_\ms$ models the contribution to $n_\xc$ on top of $n_\text{ref}$ as
\begin{equation}
\label{eq:sources}
    n _\ms (\rr) = \sum _{k = 1} ^{N_1} q_k \rho_k(\rr) - \sum _{k=1} ^{N_2} \vb p_k \cdot \grad \rho _k(\rr) \; ,
\end{equation}
with charges $q_k$ and dipole moments $\vb p _k$, respectively. In \cref{eq:sources}, sources $\rho _k$ are normalized Gaussians centered at $\vb a _k$ with finite widths set by exponents $\alpha _k$. Charge densities in \cref{eq:sources} source a parameterized electrostatic potential given by
\begin{equation}
\label{eq:potential-splat}
    v _\ms (\rr) = \sum _{k = 1} ^{N_1} q_k v_k(\rr) - \sum _{k=1} ^{N_2} \vb p_k \cdot \grad v_k(\rr) \; ,
\end{equation}
making up the total multipole splat potential. Each elementary multipole source $\rho _k$ generates an analytical electrostatic potential $v_k$, allowing us to model the source density and the resulting potential using the same parameters $\theta$. Precise functional forms of $\rho _k$ and $v_k$ have been listed in \cref{tab:multipole-dictionary}.

\begin{table}[t]
\centering
\begin{tabular}{S{c}|S{c}|S{c}}
    & Source $\rho$ & Potential $v$ \\
    \hline \hline
    \thead{Monopole\\splats} & ${\displaystyle q \; \left( \tfrac{\alpha}{\pi} \right) ^{\nicefrac32} e^{ -\alpha | \rr - \vb a |^2 }}$ & ${\displaystyle q \; \frac {\erf ( \sqrt{\alpha} | \rr - \vb a | )}{| \rr - \vb a |}}$ \\
    \hline
    \thead{Dipole\\splats} & ${\displaystyle -\vb p \cdot \grad \left\{ \left( \tfrac{\alpha}{\pi} \right) ^{\nicefrac32} e^{ -\alpha | \rr - \vb a |^2 } \right\}}$ & ${\displaystyle -\vb p \cdot \grad \frac {\erf ( \sqrt{\alpha} | \rr - \vb a | )}{| \rr - \vb a |}}$ \\
\end{tabular}
\caption{
    \textbf{Analytic forms of multipole splats.}
    An overview of the expressions for monopole and dipole source densities and resulting elementary Coulomb potentials.
}
\label{tab:multipole-dictionary}
\end{table}

If all individual $\rho_k$ are normalized to unity, the total charge of the $N_1$ monopoles is $Q_\ms = \sum _k q_k$. Practically, we define unconstrained weights $w_k$ and set
\begin{equation}
\label{eq:charges-params}
    q_k = w_k - \frac{1}{N_1} \sum _{j=1} ^{N_1} w_j + \frac{1-\gamma}{N_1}
\end{equation}
for a functional that uses an exact exchange fraction $\gamma$. \cref{eq:charges-params} enforces the sum rule $Q_\ms = \sum _k q_k = 1 - \gamma$ at all times, with controllable $\gamma$ built into the parameterization. Values of $w_k$ are unconstrained and can be optimized freely.

Physically, the variational potential is constrained to have the correct asymptotics by requiring the monopole term to carry a prescribed excess charge. Neutral multipole clouds model the exact KS potential. Charge can be added manually to model non-unit fractions of exact exchange, intentionally altering the asymptotic decay.

Variational parameters $\theta$ in \cref{eq:total-ansatz} contain all of the monopole and dipole positions, exponents, charges and dipole moments,
\begin{equation}
\label{eq:parameters}
    \theta = \Big\{ w_k, \alpha _k, \vb{a}_k \Big\} _{k=1} ^{N_1} \; \cup \; \Big\{ \vb p_k, \alpha _k, \vb{a}_k \Big\} _{k=1} ^{N_2} \; ,
\end{equation}
adding up to $P = 5 N_1 + 7 N_2$ variational parameters. Anisotropic Gaussians~\cite{kerbl3DGaussianSplatting2023} and optimizable quadrupoles or even higher floating multipoles could be included for additional expressivity in specific applications. We use the standard first-order gradient-based optimizer Adamax~\cite{kingmaAdamMethodStochastic2015} to numerically optimize all of the parameters simultaneously. Crucially, Adamax is a first-order optimizer and never constructs or attempts to approximate or invert the ill-conditioned parameter Hessian matrix~\cite{yangDirectMethodOptimized2002}. All calculations use a restricted spatial-orbital convention, with occupations in $\{0,1,2\}$. All CCSD(T) reference densities are computed in the same orbital basis as the inversion. Complete hyperparameter settings for every calculation reported above are given in \sref{tab:hyperparams}. See \sref{sec:details} for details.

There is no Coulomb singularity in \cref{eq:potential-splat}. Pure point monopoles and dipoles have delta-function source densities $q \delta (\rr - \vb a)$ and $-\vb p \cdot \grad \delta(\rr - \vb a)$, respectively, giving rise to singular Coulomb potentials. We bypass numerical instabilities associated with delta functions by modeling our sources as finite-width Gaussians defined in \cref{tab:multipole-dictionary}, keeping both densities and potentials analytic. Formally, we can recover the delta function as a limit $\lim _{\alpha \to \infty } \rho(\rr) = \delta (\rr - \vb a)$.

Other parameters in \cref{eq:parameters} are not constrained by physics. Regardless, we do parameterize the multipole positions $\vb a_k$ and the exponents in a way that constrains their magnitude. For details on parameterization and optimization, see \sref{sec:params}.

Setting $\sum_k q_k = 0$ is possible even for global hybrids and pure density XC functionals, forcing the optimization to converge to a smooth interpolation between the bulk and the exact $-\nicefrac{1}{r}$ asymptotic decay, instead of intentionally removing exact constraints to facilitate known functional deficiencies. We leave this form of adaptive range separation for future work.

Each monopole or dipole is a floating $s$ or $p$ orbital. As a consequence, potential matrix elements that contribute to the generalized splatting eigenvalue problem in \cref{eq:potential-basis} can be evaluated analytically. Combining \cref{eq:potential-basis} and the functional forms of multipole splats given in \cref{tab:multipole-dictionary}, we have
\begin{equation}
\label{eq:monopole-basis}
    V ^{(1)} _{\mu \nu} = \sum _k q_k \int \dd[3]{\rr} \int \dd[3]{\rr '} \frac{\chi _\mu (\rr) \chi _\nu (\rr) \rho_k (\rr ')}{|\rr - \rr '|}
\end{equation}
for the monopole term. A nearly identical expression can be written down for the dipole term, by replacing $q_k \rho _k \rightarrow -\vb p \cdot \grad \rho _k$. Both the integral in \cref{eq:monopole-basis} and its dipole equivalent can be evaluated using the standard three-center Gaussian integral routines commonly used for density fitting. No grid integration is required. However, we found that evaluating \cref{eq:potential-basis} on a numerical quadrature grid yielded almost identical accuracy in our experiments.
\subsection*{Regularization}
\label{sec:force-matching}

We use a physics-informed regularization term $R[v]$~\cite{heaton-burgessOptimizedEffectivePotentials2007, heaton-burgessOptimizedEffectivePotentials2008, jacobUnambiguousOptimizationEffective2011, kanungoExactExchangecorrelationPotentials2019} in \cref{eq:total-energy-reg,eq:total-wy-reg}. Minimizing the expression
\begin{equation}
\label{eq:force-matching}
    R[v] = \frac{1}{8 \pi} \int \dd[3]{\rr} \, \left| \grad v (\rr) - \grad v_0 (\rr) \right|^2
\end{equation}
enforces the smoothness bias by force matching~\cite{ercolessiInteratomicPotentialsFirstPrinciples1994} against a fixed reference potential $v_0$. Intuitively, adding the regularization term resolves the null space of near-degenerate OEP solutions in favor of smoother potentials.

In the case of multipole splat trial potentials, the regularization integral in \cref{eq:force-matching} is carried out against the reference potential $v_0 = v_\ext + v_\fa$, resulting in the electrostatic energy of the multipole splat contribution alone,
\begin{align}
    R[v_\theta]
    =& \frac{1}{8\pi} \int \dd[3]{\rr} |\grad v_\ms (\rr)| ^2 \\
    =& \frac12 \int \dd[3]{\rr} \int \dd[3]{\rr '} \frac{n_\ms (\rr) n_\ms (\rr')}{|\rr - \rr '|} \; . \label{eq:ms-reg}
\end{align}
The Coulomb energy in \cref{eq:ms-reg} can be evaluated analytically as well, completely bypassing grid integration of trial potentials in the entire optimization loop.

Regularization defined in \cref{eq:force-matching} is not required for numerical convergence. A small positive constant $\lambda$ is used to select a visually appealing potential between all of the degenerate solutions with almost identical energies. This near-degeneracy is shown to correspond to the null space of the KS response function in \sref{sec:regularization}, responsible for numerical problems in traditional solvers that rely on inversion. Visually, the optimization procedure produces potentials with correct large-scale physical features even at $\lambda \rightarrow 0$ (see \sref{sec:hyperparams} and \sref{fig:ablation-metrics} and \sref{fig:ablation-profiles}). From the physical point of view, regularization biases the optimization towards solutions with electric fields (or forces) $\vb E = -\grad v$ matching those of the background $v_0$.

In operator form, the total energy functional Hessian $\mathcal H$ at the optimum takes the form $\mathcal{H} = - \chi_s \chi^{-1} \chi_s$, where $\chi = \fdv{n}{v_\ext}$ is the density response operator. Therefore, the Hessian shares the null space with the KS response $\chi _s$, manifesting as flat directions at the optimum in the energy landscape $E[v]$. The effect of the additive regularizer in \cref{eq:force-matching} is to add its own positive definite Hessian $\mathcal{H}_{\text{reg}} = - \tfrac{1}{4\pi} \laplacian$ to that of the total energy or the quantum KL divergence, making the total Hessian strictly positive definite, up to constant shifts. The full derivation and additional details can be found in \sref{sec:regularization}.

\section*{Data availability}

The converged potentials, source densities, orbital coefficients and spectra supporting the findings of this study, together with the molecular geometries and reference densities used as inputs, are available upon reasonable request from the corresponding author. The Ne~I reference term values are available from the NIST Atomic Spectra Database~\cite{kramidaNISTAtomicSpectra1999}, and the GW100 reference ionization potentials from Ref.~\cite{vansettenGW100BenchmarkingG0W02015}.

\section*{Code availability}

All simulations were performed using the JAX~\cite{bradburyJAXComposableTransformations2018} library for array manipulation and automatic differentiation. Equinox~\cite{kidgerEquinoxNeuralNetworks2021} was used for model design, and Optax for optimization. Data was post-processed using NumPy~\cite{harrisArrayProgrammingNumPy2020} and SciPy~\cite{virtanenSciPy10Fundamental2020}. The plots were produced using the Matplotlib~\cite{hunterMatplotlib2DGraphics2007} library. The custom code library needed to reproduce the results in this work and explore new ones can be found in the following repository: \url{https://github.com/Matematija/multipole-splats}.

\bibliographystyle{naturemag}
\bibliography{references}

\section*{Acknowledgements}

M.M. acknowledges insightful discussions with Michael Ruggenthaler about theoretical aspects and Heiko Appel about numerical aspects. M.M. also acknowledges support from a SwissAI grant from the Swiss National Supercomputing Centre (CSCS) under project ID a164. This work was supported by the European Research Council (ERC-2024-SyG-101167294; UnMySt), the Cluster of Excellence Advanced Imaging of Matter (AIM). We acknowledge support from the Max Planck-New York City Center for Non-Equilibrium Quantum Phenomena. The Flatiron Institute is a division of the Simons Foundation.

\section*{Author contributions}

M.M. conceived the project, developed the theory, implemented the code, performed the calculations, and wrote the manuscript. A.R. and J.C. supervised the project and contributed to the manuscript. All authors discussed the results and commented on the manuscript.

\section*{Competing interests}
The authors declare no competing interests.

\end{document}


\title{Multipole splats for optimized and inverted effective potentials\\Supplementary Information}

\author{Matija Medvidović}
\email{mmedvidovic@ethz.ch}
\affiliation{Institute for Theoretical Physics, ETH Zürich, 8093 Zürich, Switzerland}

\author{Angel Rubio}
\affiliation{Max Planck Institute for the Structure and Dynamics of Matter, Luruper Chaussee 149, 22761 Hamburg, Germany}
\affiliation{Center for Computational Quantum Physics, Flatiron Institute, 162 5th Avenue, New York, NY 10010, USA}
\affiliation{Initiative for Computational Catalysis, Flatiron Institute, 162 5th Avenue, New York, NY 10010, USA}

\author{Juan Carrasquilla}
\affiliation{Institute for Theoretical Physics, ETH Zürich, 8093 Zürich, Switzerland}

\date{September 22, 2026}

\maketitle

\suppnote{Exact gradients}
\label{sec:gradients}

\subsection*{Optimized effective potentials}

In the main text, we minimize the total energy functional $E[\phi]$ as a functional of the orbitals such that the orbitals are compatible with a single-particle multiplicative potential. We define our trial orbitals $\phi = \phi[v]$, thinking of these orbitals as functionals of a trial single-particle potential $v$. In other words, the orbitals solve
\begin{equation}
\label{eq:single-particle-eq}
    \left(-\tfrac12 \laplacian + v(\rr) \right) \phi _i (\rr) = \epsilon _i \, \phi _i (\rr) \; ,
\end{equation}
for some $v$. Following the Kohn-Sham (KS) recipe, we seek to find the $v$ that minimizes the total density functional theory (DFT) energy $E$ as a functional of the trial potential $v$, $v _\oep  = \argmin _{v} E[\phi[v]]$.

\cref{eq:single-particle-eq} is not the Kohn-Sham equation we are trying to solve but is simply a defining equation for the trial orbitals. We explicitly parameterize the potential $v = v_\theta$, requiring us to carry the gradients through the solution of \cref{eq:single-particle-eq} during gradient-based optimization. Assuming real orbitals $\phi_i (\rr) \in \mathbbm{R}$, and using the chain rule, we expand
\begin{equation}
\label{eq:fdv-chain-rule}
    \pdv{E}{\theta _a} =
    \sum _i \int \dd[3]{\rr} \fdv{E [\phi]}{\phi ^* _i (\rr)} \pdv{\phi ^* _i (\rr)}{\theta _a} \; + \; \text{c.c.} =
    \sum _i \int \dd[3]{\rr} \int \dd[3]{\rr'} \fdv{E [\phi]}{\phi ^* _i (\rr)} \fdv{\phi ^* _i (\rr)}{v (\rr')} \pdv{v(\rr')}{\theta _a} \; + \; \text{c.c.} \; .
\end{equation}
In \cref{eq:fdv-chain-rule}, the chain rule application results in the product of three derivative factors. The first term results in the generally orbital-dependent DFT single-particle Hamiltonian. As an example, for hybrid functionals with an exact-exchange fraction $\gamma$, we have
\begin{equation}
    \fdv{E [\phi]}{\phi ^* _i (\rr)} = ( \Hat{\vb{H}} \, \phi ) _i (\rr) =
    \left[ -\tfrac{1}{2} \laplacian + v_\ext (\rr) + v_H (\rr) + v_\xc ^{(\text{loc})} (\rr) \right] \phi _i (\rr) - \gamma \sum _j K _{i j} (\rr) \phi _j (\rr) \; ,
\end{equation}
where $\Hat{ \vb{H} } = \Hat{ \vb{H} } [\phi] $ is the effective generalized Kohn-Sham (GKS) Hamiltonian operator, comprised of the external potential $v_\ext$, the local exchange-correlation (XC) contribution $v_\xc ^{(\text{loc})}$, and the standard direct (Coulomb or Hartree) term and exchange (Fock) terms,
\begin{equation}
    v_H (\rr) = \int \dd[3]{\rr'} \frac{n(\rr')}{|\rr - \rr'|}
    \; , \qquad
    K _{i j} (\rr) = \int \dd[3]{\rr'} \frac{ \phi _i (\rr ') \phi ^* _j (\rr') }{|\rr - \rr'|} \; .
\end{equation}
The orbital response factor $\fdv{\phi}{v}$ is given by first-order perturbation theory:
\begin{equation}
    \fdv{\phi _i (\rr)}{v (\rr ')} = \sum _{j \neq i} \frac{\phi _j (\rr) \phi ^* _j (\rr ') }{\epsilon _i - \epsilon _j} \phi _i (\rr ')
\end{equation}
We arrive at the final expression
\begin{align}
\label{eq:oep-gradient}
    \pdv{E}{\theta _a} =& \sum _i \sum _{j \neq i} \int \dd[3]{\rr} \int \dd[3]{\rr'} ( \Hat{\vb{H}} \, \phi ) _i (\rr) \frac{\phi ^* _j (\rr) \phi _j (\rr ') }{\epsilon _i - \epsilon _j} \phi ^* _i (\rr ') \pdv{v(\rr')}{\theta _a} + \; \text{c.c.} \\
    =& \sum _i \sum _{j \neq i} \left[ \int \dd[3]{\rr} \phi ^* _j (\rr) \Hat{\vb{H}} \; \phi _i (\rr) \right] \frac{1}{\epsilon _i - \epsilon _j} \left[ \int \dd[3]{\rr'} \phi ^* _i (\rr ') \pdv{v(\rr')}{\theta _a} \phi _j (\rr ') \right] + \; \text{c.c.} \\
    =& \sum _i \sum _{j \neq i} \frac{H _{i j} G _{i j}}{\epsilon _i - \epsilon _j} + \; \text{c.c.}
\end{align}
where $H_{i j} = \bra{\phi _i} \vb H \ket{\phi _j}$ are the GKS Hamiltonian matrix elements in the optimized effective potential (OEP) orbital basis and
\begin{equation}
    G ^a _{i j} = \int \dd[3]{\rr'} \phi ^* _i (\rr ') \pdv{v(\rr')}{\theta _a} \phi _j (\rr ') = \bra{\phi _i} \Hat{\vb G} ^a \ket{\phi _j}
\end{equation}
can be understood as matrix elements of a potential gradient operator. The gradient expression in \cref{eq:oep-gradient} has several advantages over traditional approaches. Firstly, it is numerically stable to evaluate unless $\epsilon _i \approx \epsilon _j$ in the denominator. Therefore, any divergences are directly caused by degeneracies in the MO spectrum. To stabilize the optimization in such cases, we intervene in the gradient computation in \cref{eq:oep-gradient} with
\begin{equation}
    \frac{1}{\epsilon _i - \epsilon _j}
    \quad \mapsto \quad
    \begin{cases}
       (\epsilon _i - \epsilon _j)^{-1}, & \text{if } |\epsilon _i - \epsilon _j| > \delta \\
       0,                                & \text{otherwise.}
    \end{cases}
\end{equation}
We set $\delta = 10^{-7} \; \text{Ha}$ throughout. Any remaining optimization instability is removed by {gradient clipping}~\cite{pascanuDifficultyTrainingRecurrent2012, zhangWhyGradientClipping2019a}. Gradient clipping is considered standard numerical hygiene in artificial intelligence (AI) and Gaussian splat optimization in computer vision. In fact, our numerical tests indicate that gradient clipping is sufficient for reliable optimization at $\delta = 0$.

\subsection*{Inverse Kohn-Sham}

This subsection is devoted to relating the quantum Kullback-Leibler (KL) divergence (or the von Neumann relative entropy) loss to the well-known Wu-Yang (WY) functional \cite{wuDirectOptimizationMethod2003}. This derivation is presented in support of the point of view that the inverse Kohn-Sham (IKS) problem is fundamentally a supervised Hamiltonian learning problem. The goal of this learning task is to find the best single-particle potential $v$ by minimizing the quantum KL divergence
\begin{equation}
\label{eq:qkl-appendix}
    \mathcal{D} _\text{KL} (\Hat \rho _0 \parallel \Hat \rho _\beta[v]) = - \Tr{ \Hat \rho _0 \left(\ln \Hat \rho _\beta[v] - \ln \Hat \rho _0 \right)}
\end{equation}
between the trial $\Hat \rho _\beta[v]$ and reference $\Hat \rho _0$ density operators. Both are normalized states on the fermionic Fock space.
The trial state can be reduced to the one-particle density matrix $\Gamma [v] = \sum _k f_k [v] \ketbra*{\phi _k[v]}{\phi _k[v]}$ whose diagonal gives the trial density $n[v]$. The orbitals are defined as the eigenstates of $\Hat H[v]$, represented in second quantization as
\begin{equation}
    \Hat H [v] = \int \dd[3]{\rr} \; \Hat \Psi ^\dagger (\rr) \left[ -\tfrac12 \laplacian + v (\rr) \right] \Hat \Psi (\rr)
    \qquad \text{where} \qquad
    \left\{ \Hat \Psi (\rr), \Hat \Psi ^\dagger (\rr ') \right\} = \delta (\rr - \rr')
\end{equation}
are the fermionic field operators. We set up our expressions in the grand-canonical ensemble, at finite inverse temperature $\beta$ and chemical potential $\mu$, with the intention of taking the limit $\beta \rightarrow \infty$ as the final step. In the finite-temperature extension of DFT~\cite{merminThermalPropertiesInhomogeneous1965a}, we have
\begin{equation}
    \Hat \rho _\beta[v] = \frac{e^{-\beta (\Hat H[v] - \mu \Hat N)}}{Z [v; \beta, \mu]}
    \qquad \text{and} \qquad
    Z [v; \beta, \mu] = \Tr e^{-\beta (\Hat H[v] - \mu \Hat N)} = e^{- \beta \, \Omega [v; \beta, \mu]}
\end{equation}
defining the grand-canonical potential $\Omega = E - T S - \mu N$ and the particle number operator $\Hat N = \int \dd[3]{\rr} \Hat \Psi ^\dagger (\rr) \Hat \Psi (\rr)$. This form of the density operator is the exponential family referred to in the main text, with $-\beta v$ the natural parameter conjugate to the density $n$. Defining
\begin{equation}
	N_0 = \Tr{\Hat \rho _0 \Hat N}
	\qquad \text{and} \qquad
	N_\beta[v] = \Tr{\Hat \rho _\beta[v] \Hat N}
\end{equation}
\cref{eq:qkl-appendix} expands to the familiar free energy difference
\begin{equation}
\label{eq:finite-T-KL}
    \mathcal{D} _\text{KL} (\Hat \rho _0 \parallel \Hat \rho _\beta[v]) =
    \; \beta \left[ \Tr{\Hat \rho _0 \Hat H[v] } - \Tr{\Hat \rho _\beta[v] \Hat H[v]} - \mu \left( N_0 - N_\beta[v] \right) \right] + S[\Hat \rho _\beta[v]] - S[\Hat \rho _0] \; .
\end{equation}
We restrict the reference and trial states to the same expected particle number, $N_0 = N_\beta[v]$, as in the fixed-$N$ KS problem. The chemical-potential term then vanishes. Although the relative entropy between distinct pure states may diverge as $\beta \rightarrow \infty$, its scaled limit is finite. If we decompose the trial Hamiltonian into kinetic and potential terms, we obtain
\begin{equation}
    \lim _{\beta \rightarrow \infty} \beta ^{-1} \mathcal{D} _\text{KL} (\Hat \rho _0 \parallel \Hat \rho _\beta[v]) = T_0 - T _s [\phi[v]] + \int \dd[3]{\rr} \; v(\rr) \left( n_0 (\rr) - n[v] (\rr)\right) \; ,
\end{equation}
where $T_0 = \Tr{\Hat \rho _0 \Hat T}$ is the kinetic energy of the reference state. This term does not contain any optimizable parameters and acts as an additive shift that vanishes under gradient optimization. If we drop the irrelevant constant, the resulting expression is precisely the negative of the well-known Wu-Yang functional \cite{wuDirectOptimizationMethod2003}
\begin{equation}
    W[v] = T_s [\phi[v]] + \int \dd[3]{\rr} \; v(\rr) \left( n[v] (\rr) - n_0 (\rr) \right) \; ,
\end{equation}
which attains its maximal value for the exact IKS potential $v$. Therefore,
\begin{equation}
    \argmin _v \lim _{\beta \rightarrow \infty} \beta ^{-1} \mathcal{D} _\text{KL} (\Hat \rho _0 \parallel \Hat \rho _\beta[v])  = \argmax _v W[v] \; .
\end{equation}
The connection between the quantum KL divergence and the WY functional can also be exploited to generalize the WY functional to finite-temperature density functional theory~\cite{merminThermalPropertiesInhomogeneous1965a}. At finite $\beta$, the potential-dependent part of $-\beta^{-1}\mathcal D_\text{KL}$ gives
\begin{equation}
    W_\beta [v] = E_s [f[v], \phi[v]] - T S[f[v]] - \int \dd[3]{\rr} \; v(\rr) n_0 (\rr) \; ,
\end{equation}
where $E_s [f[v], \phi[v]] - T S[f[v]]$ is the KS free energy, $W_\beta[v]$ reduces to $W[v]$ as $\beta \rightarrow \infty$, and
\begin{equation}
    S[f] = - \sum _k \left( f_k \ln f_k + (1-f_k) \ln (1-f_k) \right)
    \qquad \text{with} \quad
    f_k [v] = \frac{1}{1 + e^{\beta ( \epsilon _k [v] - \mu )}} \; .
\end{equation}

To complete the optimization loop, we need functional gradients $\fdv{W}{v}$ because they contribute to the energy gradients through $\pdv{W[v]}{\theta _a} = \fdv{W[v]}{v} \cdot \pdv{v}{\theta _a}$. Rewriting the WY functional using the total energy of the KS system as the sum of energies corresponding to occupied KS orbitals, we have
\begin{equation}
    T_s [\phi] + \int \dd[3]{\rr} v(\rr) n(\rr) = \sum _{i=1} ^N \epsilon _i
    \quad \Rightarrow \quad
    W[v] = \sum _{i=1} ^N \epsilon _i [v] - \int \dd[3]{\rr} \; v(\rr)  n_0 (\rr) \; ,
\end{equation}
ready to exploit the perturbation theory result $\fdv{\epsilon _k}{v} = |\phi _k (\cdot)| ^2$. From there, by direct computation, we have
\begin{equation}
    W[v] = \sum _i \epsilon _i [v] - \int \dd[3]{\rr} \; v(\rr)  n_0 (\rr)
    \quad \Rightarrow \quad
    \fdv{W[v]}{v} = \sum _k \left| \phi _k (\rr) \right| ^2 - n_0 (\rr) = n(\rr) - n_0 (\rr) \; .
\end{equation}
The final gradient expression establishes two major advantages of the WY functional: it is concave with respect to the trial potential and the derivatives are simply given as local density residuals, bypassing the computation of excited states completely. Consequently, its negative defines a convex loss.

\suppnote{Regularization and null spaces}
\label{sec:regularization}

The proposed regularization function in the main text can be generalized to include a spatially dependent positive weight $w (\rr) > 0$,
\begin{equation}
\label{eq:regularization-appendix}
    R[v] = \frac{1}{8 \pi} \int \dd[3]{\rr} \, w (\rr) \, \left| \grad v (\rr) - \grad v_0 (\rr) \right| ^2 \; ,
\end{equation}
to regularize oscillations in some spatial regions more than others. In preparing this manuscript, we tried many different functional forms for $w$, but none of them showed a decisive advantage over $w = \text{const.}$ for all considered systems. In all cases, the regularization in \cref{eq:regularization-appendix} is designed to select smooth potentials. This section is devoted to quantifying that statement. Let us assume for the moment that the optimization is carried out in functional space, free from finite parameterization artifacts. We restrict the analysis to a fixed-particle-number, $v$-representable branch with a non-degenerate ground state, and assume that the implicit density functional $E[n]$ is locally twice differentiable. Inverse response operators are understood on the particle-conserving range after removing the constant-potential mode. Denoting the constrained total energy functional $E[v] = E[\phi[v]]$, we construct a local quadratic approximation of the energy landscape:
\begin{equation}
\label{eq:energy-quadratic}
    E[v + \delta v] = E[v] + \int \dd[3]{\rr} \fdv{E [v]}{v (\rr)} \delta v (\rr) + \frac12 \int \dd[3]{\rr} \int \dd[3]{\rr '} \delta v (\rr) \frac{ \delta ^2 E [v]}{\delta v (\rr) \delta v (\rr ')}  \delta v (\rr ') + \mathcal{O}(\delta v ^3) \; .
\end{equation}
At convergence, the linear term in \cref{eq:energy-quadratic} vanishes because the exact OEP satisfies $\fdv{E}{v} = 0$ by definition. Therefore, the stability is determined by the spectrum of the functional Hessian operator $\mathcal H$. Using $\delta ^2 T_s / \delta n^2 = -\chi_s^{-1}$ on the restricted response space, we obtain
\begin{equation}
\label{eq:hessian}
    \mathcal H (\rr, \rr ') = \frac{ \delta ^2 E [v]}{\delta v (\rr) \delta v (\rr ')} = -\chi _s (\rr, \rr') + \int \dd[3]{\rr _1} \int \dd[3]{\rr _2} \chi _s (\rr, \rr_1) \mathcal K _\text{Hxc} (\rr_1, \rr_2) \chi _s (\rr_2, \rr ') \; ,
\end{equation}
or $\mathcal H = -\chi _s + \chi _s \mathcal{K}_\text{Hxc} \chi _s$ in the terse operator notation, with the Kohn-Sham response defined as $\chi _s = \fdv{n}{v}$, and the Hartree-exchange-correlation (HXC) kernel as
\begin{equation}
    \mathcal K _\text{Hxc} (\rr, \rr ') = \frac{1}{|\rr - \rr '|} + \frac{ \delta ^2 E _\xc}{\delta n (\rr) \delta n (\rr ')} \; .
\end{equation}
Because $v = v_\ext + v_\text{Hxc}$ at convergence, we have a clean connection with the total response $\chi = \fdv{n}{v_\ext}$
\begin{equation}
    \mathcal{K}_\text{Hxc} = \chi _s ^{-1} - \chi ^{-1}
    \qquad \Rightarrow \qquad
    \mathcal H = -\chi _s \chi ^{-1} \chi _s \; ,
\end{equation}
where we have used \cref{eq:hessian}. At a stable ground state, $\chi$ is negative definite on the restricted response space. Therefore, for any potential perturbation $\delta v$,
\begin{equation}
    \delta v^\top \mathcal H \delta v = - (\chi_s \delta v)^\top \chi^{-1} (\chi_s \delta v) \geq 0,
\end{equation}
with equality holding if and only if $\chi_s \delta v = 0$. Hence, $\mathcal H$ and $\chi_s$ share the same null space under the assumptions stated above. In the complete functional formulation, the static response has only the constant-potential zero mode for a non-degenerate ground state, consistent with the Hohenberg-Kohn uniqueness theorem~\cite{hohenbergInhomogeneousElectronGas1964}. In a finite orbital basis, additional potential perturbations can be invisible to the finite orbital-product space~\cite{gorlingRelationExchangeonlyOptimized2008}. If the tangent space of $v_\theta$ contains such perturbations, the null space is artificially expanded to include highly oscillatory functions $\varphi$ for which $\chi_s \varphi = \mathcal H \varphi = 0$.

Any such parameterization, linear or nonlinear, induces a tangent space basis expansion of the form
\begin{equation}
\label{eq:potential-tangent-space}
    \delta v (\rr) \quad \mapsto \quad \delta v _\theta (\rr) = \sum _\mu \pdv{v_\theta (\rr)}{\theta ^\mu} \delta \theta ^\mu \; ,
\end{equation}
where $\varphi _\mu (\rr) = \pdv{v_\theta (\rr)}{\theta ^\mu}$ are thought of as basis functions indexed by $\mu$ and $\delta \theta ^\mu$ as the expansion coefficients. The expansion in \cref{eq:potential-tangent-space} combines with the Hessian \cref{eq:energy-quadratic} to project the Hessian operator into the tangent space basis at convergence or self-consistency
\begin{equation}
    \mathcal{H} \quad \mapsto \quad \mathcal{H} _{\mu \nu} = \varphi _\mu ^\top \mathcal H \varphi _\nu = - ( \chi _s \varphi _\mu ) ^\top \chi ^{-1} ( \chi _s \varphi _\nu ) \; ,
\end{equation}
in operator notation, deployed to avoid quadruple integrals. For a nonlinear parameterization, the exact parameter Hessian contains an additional curvature term,
\begin{equation}
\label{eq:param-hessian}
    \pdv[2]{E}{\theta^\mu}{\theta^\nu} = \mathcal H_{\mu\nu} + \int \dd[3]{\rr} \,\fdv{E}{v(\rr)} \pdv[2]{v_\theta(\rr)}{\theta^\mu}{\theta^\nu} \; .
\end{equation}
The extra term in \cref{eq:param-hessian} vanishes at the functional stationary point assumed above. Away from that point, \cref{eq:potential-tangent-space} gives the pullback of the functional Hessian rather than the complete parameter Hessian. As noted in the main text, we optimize a regularized energy expression. Therefore, the Hessian operator introduced by \cref{eq:energy-quadratic} needs to be supplemented with the Hessian of \cref{eq:regularization-appendix},
\begin{equation}
    \mathcal{H} ^{(R)} (\rr, \rr ') = \frac{\delta ^2 R[v]}{\delta v (\rr) \delta v (\rr ')} = - \frac{1}{4 \pi} \div \left( w (\rr) \, \grad \delta (\rr - \rr ') \right) \; ,
\end{equation}
yielding its finite-basis form
\begin{equation}
\label{eq:reg-hessian}
    \mathcal{H} ^{(R)} _{\mu \nu} = \frac{1}{4 \pi} \int \dd[3]{\rr} w (\rr) \; \grad \varphi _\mu (\rr) \cdot \grad \varphi _\nu (\rr) \; ,
\end{equation}
after simplification by partial integration. If $w (\rr)$ is positive everywhere, its quadratic form is strictly positive for every non-constant potential perturbation. The matrix in \cref{eq:reg-hessian} is therefore positive definite when the tangent functions $\{\varphi_\mu\}$ are linearly independent, up to constant shifts. The regularized total energy Hessian operator at convergence reads
\begin{equation}
    \mathcal{H} _\text{tot} = - \int \dd[3]{\rr _1} \int \dd[3]{\rr _2} \; \chi _s (\rr, \rr _1) \chi ^{-1} (\rr_1, \rr _2) \chi _s (\rr_2, \rr ') - \frac{\lambda}{4 \pi} \div \left( w (\rr) \, \grad \delta (\rr - \rr ') \right) \; .
\end{equation}
At a local minimum, the unregularized Hessian is positive semidefinite. The total Hessian is therefore positive definite on the gauge-fixed potential tangent space as long as $\lambda w (\rr) > 0$ and the parameterization contains no redundant tangent directions.

\suppnote{Multipole splat parameterization}
\label{sec:params}

As noted in the main text, optimizable parameters $\theta$ consist of individual multipole parameters -- monopole charges $q_k$, dipole moments $\vb p_k$ as well as positions $\vb a _k$ and exponents $\alpha _k$. For completeness, the total variational potential can be written down as an electrostatic Coulomb potential
\begin{equation}
    v_\theta (\rr) = \int \dd[3]{\rr'} \, \frac{\rho_\theta (\rr')}{|\rr - \rr '|}
\end{equation}
sourced by a charge density
\begin{equation}
\label{eq:total-source}
    \rho _\theta (\rr) = \rho _\ext (\rr) + \tfrac{N-1}{N} n_\text{ref} (\rr) + \sum _k q_k \rho _k (\rr) - \sum _k \vb p_k \cdot \grad \rho_k (\rr) \; ,
\end{equation}
where $n_\text{ref}$ is the reference (Hartree-Fock) density and
\begin{equation}
    \rho _\ext (\rr) = -\sum _a Z_a \delta(\rr - \RR_a)
    \; ; \qquad
    \rho _k (\rr) = \left( \frac{\alpha _k}{\pi} \right)^{\nicefrac{3}{2}} e^{-\alpha _k |\rr - \vb a _k|^2} \; .
\end{equation}
Terms in \cref{eq:total-source} separate the analytically known static nuclear contribution in the first term from the precomputed Fermi-Amaldi (FA) background in the second term. The source density in \cref{eq:total-source} generates the Coulomb potential
\begin{equation}
\label{eq:total-potential}
    v _\theta (\rr) = v _\ext (\rr) + v_\text{FA} (\rr) + \sum _k q_k v_k (\rr) - \sum _k \vb p_k \cdot \grad v_k (\rr) \; ,
\end{equation}
where
\begin{equation}
    v _\ext (\rr) = -\sum _a \frac{Z_a}{|\rr - \RR _a|}
    \; , \qquad
    v_\text{FA} (\rr) = \frac{N-1}{N} \int \dd[3]{\rr'} \, \frac{n _\text{ref} (\rr ')}{|\rr - \rr '|}
    \qquad \text{and} \qquad
    v_k (\rr) = \frac{ \erf (\sqrt{\alpha _k} |\rr - \vb a _k|) }{|\rr - \vb a _k|} \; .
\end{equation}
Analytically transitioning from \cref{eq:total-source} to \cref{eq:total-potential} is made possible by convenient integral identities,
\begin{equation}
    \left( \frac{\alpha _k}{\pi} \right)^{\nicefrac{3}{2}} \int \dd[3]{\rr'} \, \frac{e^{-\alpha _k |\rr ' - \vb a _k|^2}}{|\rr - \rr '|} = \frac{ \erf (\sqrt{\alpha _k} |\rr - \vb a _k|) }{|\rr - \vb a _k|}
    \qquad \text{and} \qquad
    \int \dd[3]{\rr'} \, \frac{ \grad ' \rho (\rr')}{|\rr - \rr '|} = \grad \int \dd[3]{\rr'} \, \frac{ \rho (\rr')}{|\rr - \rr '|} \; ,
\end{equation}
or by making use of Gauss' law for the spherically-symmetric Gaussian charge distribution $\rho _k$.

The only physical constraint on the parameters is the restriction on the total charge to lock in the asymptotic decay of both the source density and the potential. As noted in the main text, for $N_1$ monopoles, we set
\begin{equation}
\label{eq:charges-params}
    q_k = w_k - \frac{1}{N_1} \sum _{j=1} ^{N_1} w_j + \frac{1-\gamma}{N_1}
\end{equation}
ensuring that $\sum_k q_k = 1-\gamma$ for a hybrid functional with a fraction of exact exchange equal to $\gamma$. Parameters $w_k \in \mathbbm{R}$ are completely free and yield a valid monopole cloud with the correct total charge for all values during optimization. Dipole moments $\vb p$ are directly optimized as they do not contribute to the total charge and cannot ruin the asymptotics.

Multipole positions and Gaussian exponents are parameterized in a way that restricts their values in the environment of nuclei at coordinates $\RR _a$ with charges $Z_a$. Each multipole position $\vb a$ is modeled as
\begin{equation}
\label{eq:metaball}
    \vb a = \vb A + R_T \, \tanh \left( \frac{|\vb b - \vb A|}{R_T} \right) \; \frac{\vb b - \vb A}{|\vb b - \vb A|}
\end{equation}
where $\vb A _k = \sum _a \mathcal{A} _{k a} \RR _a$ are the individual splat anchor points constructed as a convex linear combination using the affinity matrix
\begin{equation}
\label{eq:affinity}
	\mathcal{A} _{k a} = \frac{e^{-\omega \flatfrac{|\vb b _k - \RR _a|^2}{\Lambda _a ^2} }}{\sum _c e^{-\omega \flatfrac{|\vb b_k - \RR _c|^2}{\Lambda _c ^2} }}
\end{equation}
and unconstrained vector $\vb b$. Atom weights in \cref{eq:affinity} are defined by the distance to each nucleus in units of the furthest-neighbor interatomic distance $\Lambda _a = \max _{b \neq a} |\RR _a - \RR_b|$ and the positive dimensionless snap exponent $\omega = 0.8$.

The point of \cref{eq:metaball} is to prevent multipoles from drifting too far away from the nuclei, spoiling the asymptotic behavior in practice. During optimization, they are allowed to exit the interatomic region at most one tube radius $R_T$ away from its anchor point $\vb A$. We set $R_T = \eta d_\text{max}$ where $\eta = 0.8$ and
\begin{equation}
d_\text{max} =
\begin{cases}
   \max _{a, b} |\RR _a - \RR _b|, & \text{if } N_\text{atm} > 1 \\
   R_\text{vdW},                   & \text{if } N_\text{atm} = 1
\end{cases}
\end{equation}
is the largest geometric length scale in the molecule with $N_\text{atm}$ atoms or the van der Waals radius $R_\text{vdW}$ of a single atom.

Gaussian exponents $\alpha _k$ need to stay positive and quickly adapt their values between different orders of magnitude. Therefore, we adopt the following parameterization:
\begin{equation}
\label{eq:exponent-params}
    \alpha = \alpha _\text{min} \times \left(  \frac{\alpha _\text{max}}{\alpha _\text{min}}\right) ^{s(x)}
    \qquad \text{where} \qquad
    s(x) = \frac{1}{1 + e^{-x}} \; .
\end{equation}
By optimizing the unconstrained free variable $x$ instead of $\alpha$ directly, we recover numerical stability and ensure that all exponents satisfy $\alpha _\text{min} < \alpha < \alpha _\text{max}$ for any value $x \in \mathbbm{R}$. The per-splat exponent bounds are automatically assigned at initialization as $\alpha _\text{min} = (2 L_\text{max}^2)^{-1}$ and $\alpha _\text{max} = 4 Z_\text{eff} ^2 $, where $Z_{\text{eff}, k} = \sum_a \mathcal{A} _{k a} Z_a$ and
\begin{equation}
	L_\text{max} = \max \left( R_\text{vdW}, 2 \sqrt{\ell _x ^2 + \ell _y ^2 + \ell _z ^2} \right) \; ,
\end{equation}
with $\ell ^2 _i = \Var_p \RR_{a, i}$ being the charge-weighted variance of nuclear positions with respect to the normalized distribution $p_a \propto Z_a$. After initialization, the exponent bounds are frozen and treated as constants during optimization.

Physical initialization of multipole positions reduces the number of optimization steps that need to be taken. We exploit access to spatial profiles of converged reference orbitals (Hartree-Fock in all studied systems) to construct localized orbitals. Using the Boys localization procedure, we construct localized $\varphi _k$ as the minimizer of the variance of the position operator $\Hat{\vb r}$,
\begin{equation}
\label{eq:boys}
    \mathcal L _\text{Boys} = \sum _l \left[ \bra{\varphi _l} \Hat{\vb r} ^2 \ket{\varphi _l} - \bra{\varphi _l} \Hat{\vb r} \ket{\varphi _l} ^2  \right] \; .
\end{equation}
Optimizing \cref{eq:boys} ensures that the localized orbitals have minimal spatial extent. Individual real-space means $\bm \mu _l =  \bra{\varphi _l} \Hat{\vb r} \ket{\varphi _l}$ and variances $\sigma _l ^2 = \bra{\varphi _l} \Hat{\vb r} ^2 \ket{\varphi _l} - \bra{\varphi _l} \Hat{\vb r} \ket{\varphi _l} ^2$ are a natural byproduct of the Boys optimization. We use $\bm \mu _l$ and $\sigma _l$ to initialize the Gaussian positions $\vb a$ and exponents $\alpha$. For $N_\text{splat}$ splats and $N_\text{orb}$ states, each localized orbital is allocated $N_\text{splat} / N_\text{orb}$ splats. For a given splat $k$ assigned to orbital $l$, we set
\begin{equation}
\label{eq:params-init}
    \vb a_k = \bm \mu _l + \sigma _l \, \vb z
    \qquad \text{and} \qquad
    \alpha _k = \frac{e^{\kappa u}}{2 \sigma _l ^2} \; ;
\end{equation}
where $\kappa = 1$, $\vb z \sim \mathcal N (0, \mathbbm 1)$, and $u \sim \mathcal N(0,1)$ are sampled independently for each multipole. Initialization according to \cref{eq:params-init} ensures that positions have a normal distribution around the centers of localized orbitals while widths are distributed according to the log-normal distribution to cover several orders of magnitude around the localized orbital width.

In small systems with one or two electrons, the Boys initialization would initialize all splats in one point or a narrow region. In those cases, we default to initializing all splat positions by assigning a fixed fraction of all splats to each atom $a$, and sampling the final position from a Gaussian centered at that atom, with width set to a fixed fraction of its vdW radius $R_{\text{vdW}, a}$: $\vb a \sim \mathcal N (\RR _a, \xi ^2 R_{\text{vdW}, a} ^2 \mathbbm 1)$, with $\xi = 0.8$.

\begin{figure}[!t]
    \centering
    \includegraphics[width=\linewidth]{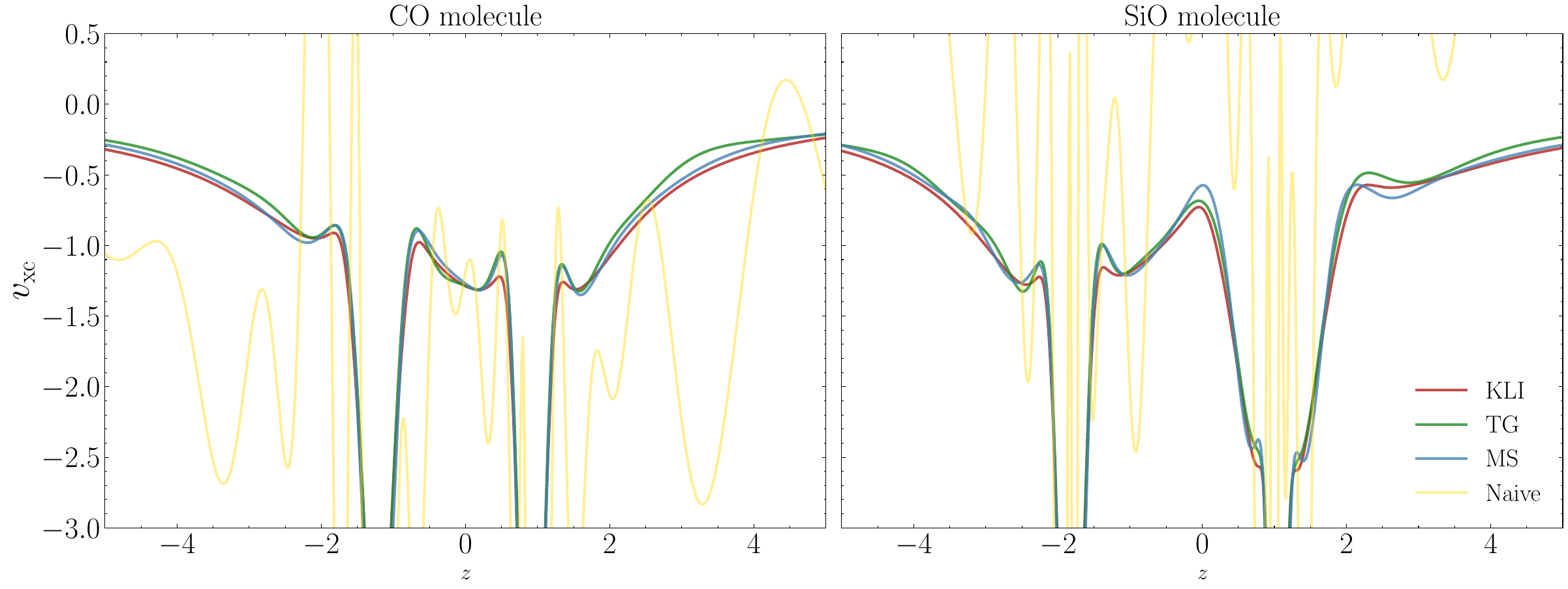}
    \caption{
        \textbf{A visual comparison of exchange potentials between finite-basis methods}.
        The Krieger-Li-Iafrate (KLI)~\cite{kriegerSystematicApproximationsOptimized1992, kriegerConstructionApplicationAccurate1992} semi-analytical OEP approximation, the basis pre-processing method of Trushin and Görling (TG)~\cite{trushinNumericallyStableOptimized2021} are plotted against multipole splats (MS). The result of a single unregularized KS response inversion starting from Hartree-Fock is also shown (naive).
        \textbf{Left}: Carbon monoxide (CO).
        \textbf{Right}: Silicon monoxide (SiO).
    }
    \label{fig:comparison}
\end{figure}

\suppnote{Comparison with other exact exchange solvers}
\label{sec:comparison}

In the traditional exact exchange (EXX) literature, the finite-basis expansion of the potential is modeled as an implicit density functional, in order to fit into the self-consistent field (SCF) framework. As noted in the main text, this density functional is evaluated by inverting the KS density response function $\chi _s = \fdv{n}{v}$, introducing numerical stability issues. However, formally, the potential can be evaluated as a solution to a Fredholm integral equation of the first kind. In the operator language, this equation can concisely be derived by careful application of the chain rule of functional derivatives,
\begin{equation}
	\fdv{E_\xc}{n} = \fdv{E_\xc}{v} \cdot \fdv{v}{n}
	\qquad \Rightarrow \qquad
	\fdv{n}{v} \cdot \fdv{E_\xc}{n} = \fdv{E_\xc}{v}
	\qquad \Leftrightarrow \qquad
	\int \dd[3]{\rr '} \; \chi _s (\rr, \rr ') v_\xc (\rr ') = \fdv{E_\xc}{v (\rr)} \; .
\end{equation}
It is natural to represent the potential $v_\xc$ in an auxiliary basis and attempt to invert the resulting matrix. When interfaced with a standard self-consistent field relaxation, one such linear solve is required per iteration. For completeness, we state the finite-basis implicit exchange functional evaluation here. For an orbital basis expansion $\ket*{\phi _k} = \sum _\mu C _{\mu k} \ket*{\chi _\mu}$ coupled with an auxiliary basis potential expansion $v_\text{x} (\rr) = \sum _\alpha v_\alpha \varphi _\alpha (\rr)$, the potential basis functions are generated by source densities $\rho_\alpha$ as $\varphi_\alpha(\rr) = \int \dd[3]{\rr'} \, \rho_\alpha(\rr') / |\rr-\rr'|$. We have the linear system $X \vb{v} = \vb{t}$ where
\begin{equation}
    X_{\alpha \beta} = \sum _{k \in \mathrm{occ}} \sum _{l \in \mathrm{virt}} D_{k l, \alpha} \frac{1}{\epsilon _k - \epsilon _l} D_{k l, \beta}
    \qquad \text{and} \qquad
    t_\alpha = - \sum _{k \in \mathrm{occ}} \sum _{l \in \mathrm{virt}} D_{k l, \alpha} \frac{1}{\epsilon _k - \epsilon _l} \bra{\phi _k} K \ket{\phi _l}
\end{equation}
with $D_{k l, \alpha}$ related to the standard three-center integrals $(\mu \nu | \alpha )$ commonly evaluated in density fitting routines
\begin{equation}
	D_{k l, \alpha} =
	\int \dd[3]{\rr} \int \dd[3]{\rr '} \frac{ \phi _k (\rr) \phi _l (\rr) \rho _\alpha (\rr')}{|\rr - \rr '|} =
	\sum _{\mu \nu} C _{\mu k} C _{\nu l} \int \dd[3]{\rr} \int \dd[3]{\rr '} \frac{ \chi _\mu (\rr) \chi _\nu (\rr) \rho _\alpha (\rr ')}{|\rr - \rr '|} =
	\sum _{\mu \nu} C _{\mu k} C _{\nu l} (\mu \nu | \alpha)
\end{equation}
and $K$ is the positive exchange operator defined in \cref{sec:gradients}, so the nonlocal exchange contribution to the Hartree-Fock Hamiltonian is $-K$.

The naive unregularized approach, simply evaluating $\vb v_\text{x} = X^{-1} \vb{t}$, is not stable numerically, as the matrix is ill-conditioned for two independent reasons:
\begin{itemize}
	\item \textbf{Finite auxiliary basis artifacts}: Finite basis expansion grows the null space of the KS from just all constant shifts of the potential to many highly oscillatory functions that are nearly orthogonal to the orbital-product space. They carry low weight in the singular value decomposition of $X$, but get amplified in the inverse matrix $X^{-1}$.
	\item \textbf{Orbital and auxiliary basis mismatch}: Coupling an insufficient auxiliary basis with a large orbital basis (or vice-versa) can result in convergence to unphysical potentials that optimize expressivity gaps instead of physical features.
\end{itemize}
We show the resulting unphysical potential for illustrative purposes on \cref{fig:comparison}, after just one exact inversion starting from the converged Hartree-Fock (HF) orbitals.

Trushin and Görling (TG)~\cite{trushinNumericallyStableOptimized2021, trushinImprovingExchangeCorrelationPotentials2025} have introduced a physically principled and numerically stable auxiliary basis preprocessing approach that mitigates both classes of problems. They eliminate basis functions that violate the target $\nicefrac{-1}{r}$ asymptotics before starting the SCF procedure, while giving recommended basis pairs and a default value for the response matrix singular value cutoff. Their approach comes in two variants: the "bare" approach described above and a version that additionally enforces the \textit{highest occupied molecular orbital (HOMO) condition}
\begin{equation}
\label{eq:homo-condition}
	\bra{\phi _\text{HOMO}} \Hat v_\text{x} \ket{\phi _\text{HOMO}} = \bra{\phi _\text{HOMO}} \Hat v_\text{x} ^\text{NL} \ket{\phi _\text{HOMO}}
\end{equation}
in preprocessing the auxiliary potential basis. In \cref{eq:homo-condition}, $\Hat v_\text{x}$ is the KS exchange potential operator and $\Hat v_\text{x} ^\text{NL}$ is the non-local (Hartree-Fock) exchange potential formed with KS orbitals. Both variants of the TG approach have allowed the auxiliary-basis OEP to be applied to a wider range of systems.

The HOMO condition is specific to EXX. It is recovered by a complete OEP solution but need not be satisfied in finite orbital and potential representations. Ref.~\cite{trushinNumericallyStableOptimized2021} enforces it explicitly during auxiliary-basis preprocessing in a variant of their approach (TG-HC), while the Krieger-Li-Iafrate (KLI) semi-analytical potential~\cite{kriegerSystematicApproximationsOptimized1992, kriegerConstructionApplicationAccurate1992} fixes the corresponding highest-occupied shift by construction. HF retains the nonlocal exchange operator and is not a local potential calculation. Multipole splat potentials instead enforce the exchange charge condition and the resulting $\nicefrac{-1}{r}$ asymptote. We do not impose \cref{eq:homo-condition} because it has no simple analogue for correlation potentials, while the present framework targets general orbital-dependent functionals. Agreement with the real-space HOMO levels is an independent benchmark for our calculations precisely because no HOMO eigenvalue or other spectral target enters the optimization. Therefore, the bare TG calculation is the closest methodological comparison.

In \cref{tab:method-comparison}, we compare the results of multipole splat direct optimization to both variants of the TG approach using the recommended \texttt{aug-cc-pwCVQZ} orbital basis. We compare the negative HOMO eigenvalues $-\epsilon_\text{HOMO}$ and total EXX energies using the numerical values reported in Ref.~\cite{trushinNumericallyStableOptimized2021}. We use atomic systems where accurate EXX reference HOMO eigenvalues are available, calculated on real-space grids by TG as well using the RELKS code~\cite{engelLocalNonlocalRelativistic1995, engelGeneralizedGradientApproximation1996, engelRelativisticOptimizedpotentialMethod1998}.

\begin{figure}[h]
    \centering
    \includegraphics[width=0.9\linewidth]{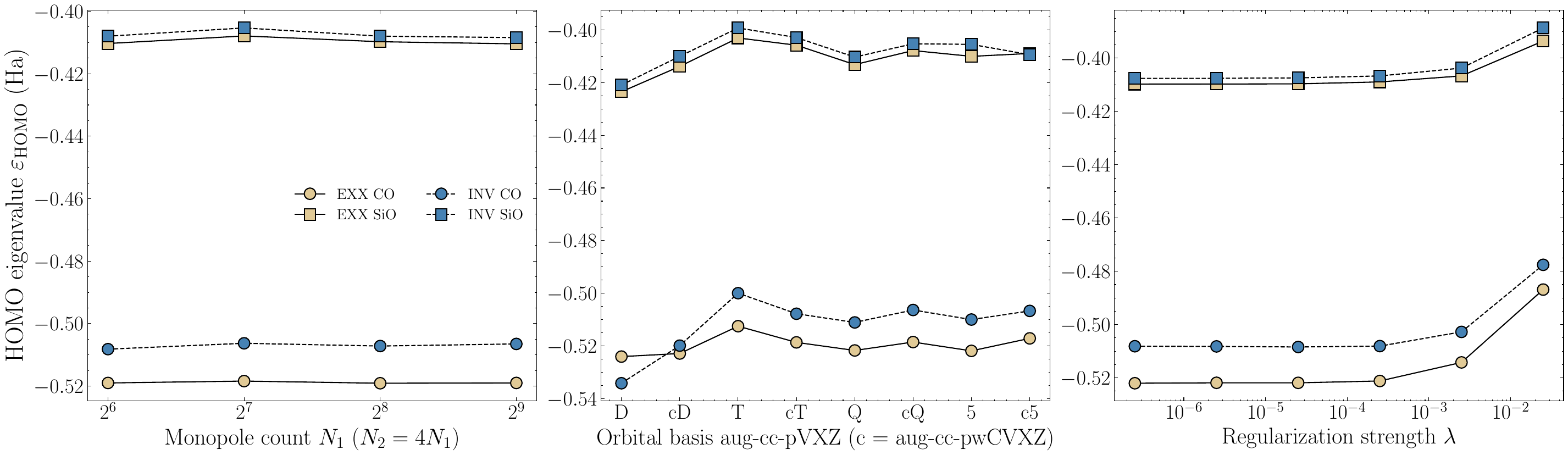}
    \hfill
    \includegraphics[width=0.9\linewidth]{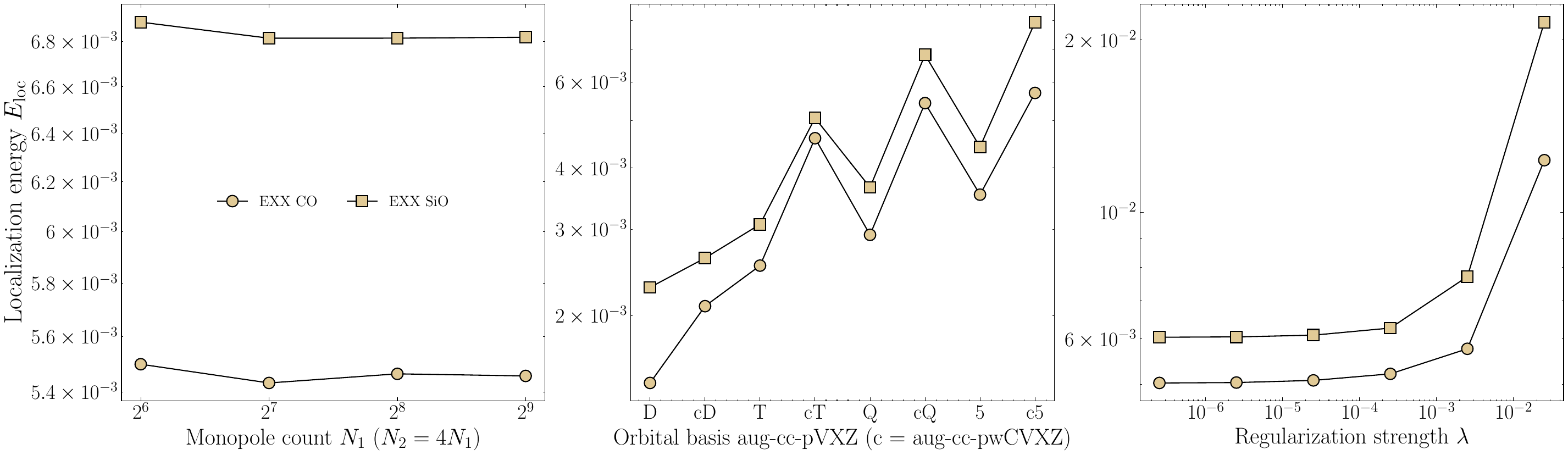}
    \hfill
    \includegraphics[width=0.9\linewidth]{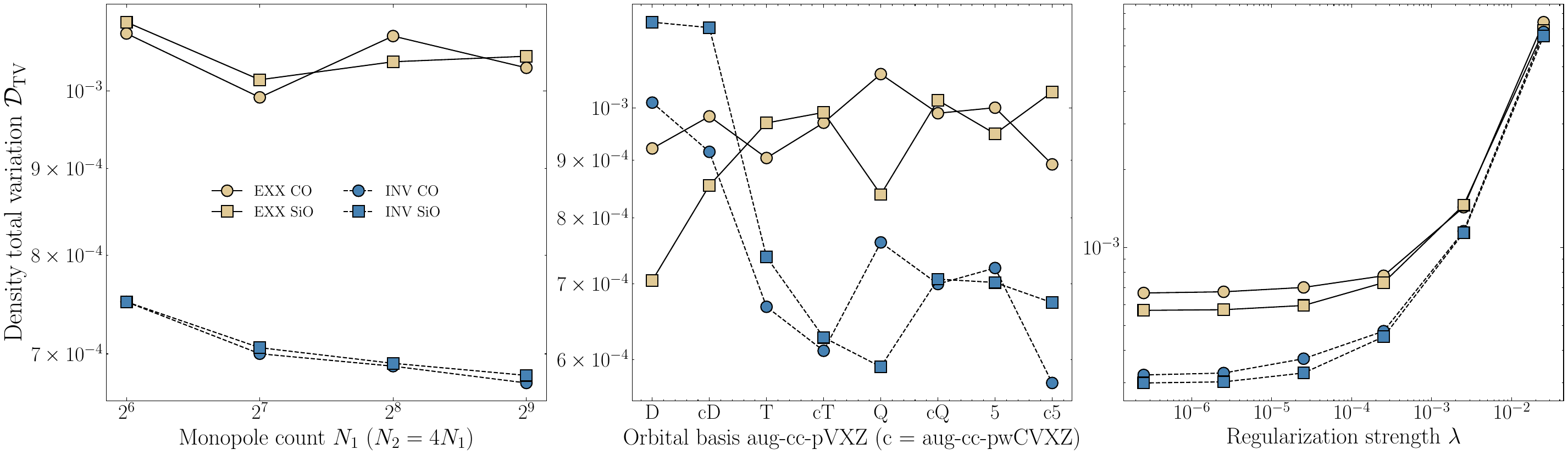}
    \caption{
        \textbf{Ablation studies for key hyperparameters in multipole splat optimization.}
        Hyperparameter ablations for CO and SiO, computed with EXX OEP (solid) and IKS inversion of a coupled-cluster singles, doubles and perturbative triples (CCSD(T)) reference density (dashed, INV).
        \textbf{Top}: the HOMO eigenvalue $\epsilon_\text{HOMO}$.
        \textbf{Middle}: the localization energy $E_\text{loc}$, defined only for the OEP.
        \textbf{Bottom}: the density total variation $\mathcal{D}_\text{TV}$.
        Each row varies one hyperparameter at a time: the monopole count $N_1$ at fixed ratio $N_2 = 4 N_1$ (left), the orbital basis (center), and the regularization strength $\lambda$ (right).
        The basis axis interleaves the \texttt{aug-cc-pVXZ} and core-valence \texttt{aug-cc-pwCVXZ} families.
    }
    \label{fig:ablation-metrics}
\end{figure}

Values sufficiently close ($\sim 10$ mHa) to the Hartree-Fock value are a necessary but not sufficient condition for retrieving the physical exchange potential. Multipole splat potentials produce energies of the correct magnitude.

Another necessary condition is an accurate HOMO eigenvalue. For the EXX model, the real-space reference values provide an independent spectral benchmark. For all five atoms with available reference values, the multipole-splat HOMO eigenvalue is closer to the reference than the bare TG result. The mean absolute deviations are $7.1$ mHa for MS and $25.3$ mHa for TG. This comparison should not be interpreted as a general ranking of the methods: TG-HC explicitly enforces the EXX HOMO condition, KLI satisfies the corresponding condition by construction, and HF retains the nonlocal exchange operator. The MS values emerge without direct HOMO enforcement.

We supplement numerical comparisons on atomic systems with a visual comparison of potential profiles in molecular systems in \cref{fig:comparison} against the KLI potential for CO and SiO molecules presented in the main text. We see that the multipole splat potentials produce smooth potentials with sharp physical features.

\begin{table}[t]
    \begin{tabular}{l|cccccc|ccccc}
        \hline
        & \multicolumn{6}{c}{$-\epsilon _\text{HOMO}$ (Ha)} & \multicolumn{5}{c}{$E_\exx$ (Ha)} \\
        \toprule
        & \textbf{MS} & TG & TG-HC & KLI & HF & Ref. & \textbf{MS} & TG & TG-HC & KLI & HF \\
        \midrule
        CO & 0.52035 & 0.48059 & 0.55359 & 0.55294 & 0.55532 & -- & -112.783355 & -112.782858 & -112.783034 & -112.781157 & -112.788791 \\
        Be & 0.30114 & 0.31896 & 0.30999 & 0.30889 & 0.30927 & 0.30923 & -14.572852 & -14.572136 & -14.571606 & -14.572374 & -14.572969 \\
        Ne & 0.83477 & 0.81089 & 0.85227 & 0.84964 & 0.85066 & 0.85046 & -128.542119 & -128.542114 & -128.541790 & -128.541517 & -128.543763 \\
        Mg & 0.24968 & 0.23946 & 0.25401 & 0.25256 & 0.25305 & 0.25302 & -199.613729 & -199.610065 & -199.610796 & -199.612663 & -199.614234 \\
        Ar & 0.58605 & 0.53414 & 0.59147 & 0.58933 & 0.59107 & 0.59040 & -526.812562 & -526.811113 & -526.811415 & -526.810031 & -526.816805 \\
        Zn & 0.28854 & 0.28558 & 0.29303 & 0.29185 & 0.29251 & 0.29280 & -1777.837773 & -1777.837880 & -1777.836633 & -1777.832169 & -1777.848027 \\
        \bottomrule
    \end{tabular}
    \caption{
        \textbf{EXX HOMO eigenvalues and total energies.}
        Negative HOMO eigenvalues $-\epsilon_\text{HOMO}$ (left) and total energies (right) of the multipole-splat EXX-OEP (MS), the auxiliary-basis EXX OEP of Trushin and Görling~\cite{trushinNumericallyStableOptimized2021} without (TG) and with (TG-HC) the HOMO condition, the Krieger-Li-Iafrate (KLI) approximation, and Hartree-Fock (HF) in the \texttt{aug-cc-pwCVQZ} orbital basis. TG and TG-HC from Table VIII in Ref.~\cite{trushinNumericallyStableOptimized2021}. The Ref. column gives real-space EXX-OEP reference values computed with RELKS and reported in the same work. TG is the closer methodological analogue of MS.
    }
    \label{tab:method-comparison}
\end{table}

\begin{figure}[t]
    \centering
    \includegraphics[width=0.9\linewidth]{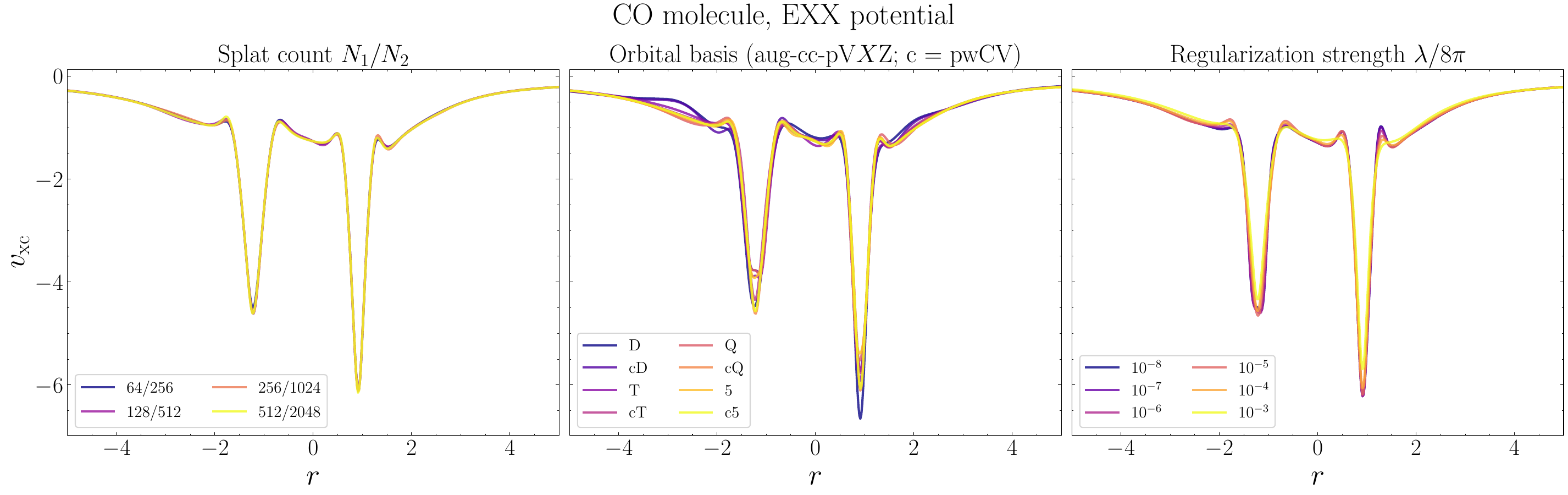}
    \hfill
    \includegraphics[width=0.9\linewidth]{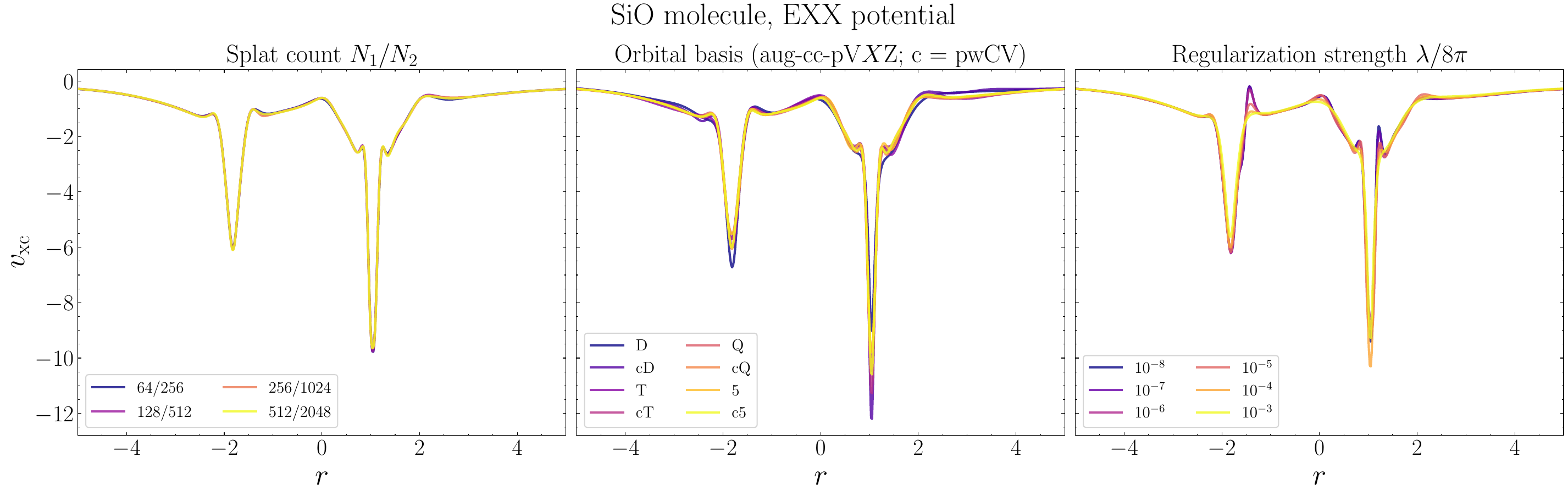}
    \hfill
    \includegraphics[width=0.9\linewidth]{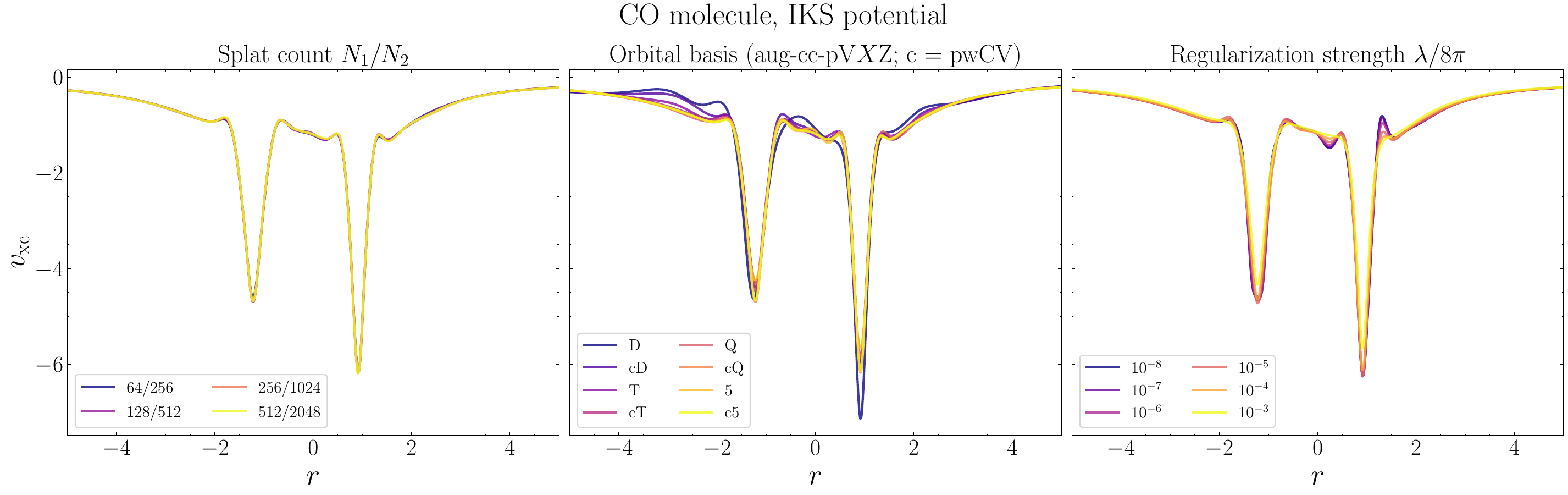}
    \hfill
    \includegraphics[width=0.9\linewidth]{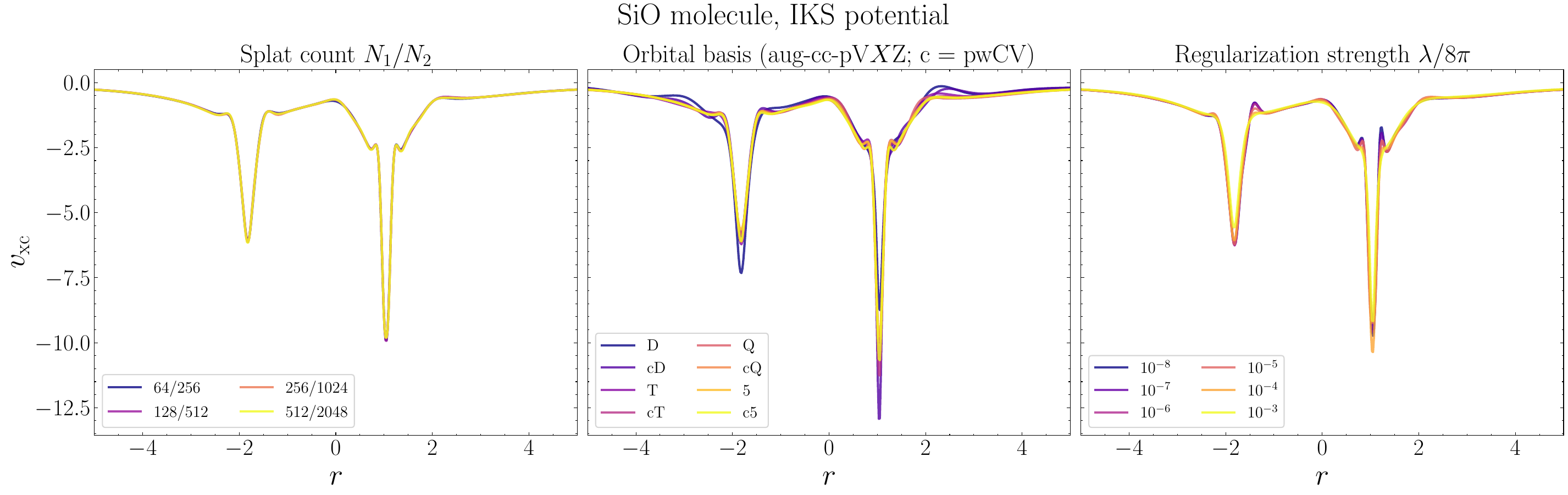}
    \caption{
        \textbf{Spatial potential profiles under ablated hyperparameters.}
        Converged XC potential profiles along the internuclear axis under the same ablations as \cref{fig:ablation-metrics}. From top to bottom: CO with EXX OEP, SiO with EXX OEP, CO with IKS, and SiO with IKS. Within each block, the splat counts $N_1/N_2$ (left), the orbital basis (center), and the regularization strength $\lambda$ (right) are varied one at a time.
    }
    \label{fig:ablation-profiles}
\end{figure}

\suppnote{Hyperparameter ablations}
\label{sec:hyperparams}

In this section, we present the results of hyperparameter ablations for the multipole splat EXX OEP and IKS. We vary the splat monopole counts $N_1$, splat dipole counts $N_2$, the size of orbital basis as X in \texttt{aug-cc-pVXZ} and the regularization parameter $\lambda$. Potentials are compared in real-space plots as well as through localization energies and total variation between OEP and reference densities.

Our main findings are:
\begin{itemize}
	\item Converged potential profiles are visually indistinguishable across a factor of eight in splat count. All three diagnostics (the HOMO eigenvalue, the localization energy, and the density total variation) are essentially flat over this range.
	\item As noted in the main text, localization energies are a non-monotonically moving target with respect to a growing orbital basis. They rise monotonically within each basis family, but adding core-valence functions raises $E_\text{loc}$ enough that its value at \texttt{cT} exceeds the one at \texttt{Q}, and \texttt{cQ} exceeds \texttt{5}. Neither $\epsilon_\text{HOMO}$ nor $\mathcal{D}_\text{TV}$ inherits this sawtooth pattern, both staying within a narrow band from triple-zeta onward. The double-zeta basis is the only one departing appreciably from the converged potential shape, underestimating the depth of the core wells.
	\item All three diagnostics display the expected elbow-like behavior with respect to the regularization parameter $\lambda$. They remain unchanged for $\lambda \lesssim 2.5 \times 10^{-4}$, degrading only once the regularizer starts to compete with the primary objective at $\lambda \approx 10^{-2}$. The Pareto-optimal $\lambda$ can be fine-tuned to this elbow for each molecule for added control over regularization effects.
	\item Lowering $\lambda$ below $\sim 2.5 \times 10^{-5}$ sharpens the core wells and intershell features slightly. However, the textbook OEP oscillations are suppressed even at lowest values of $\lambda \approx 2.5 \times 10^{-7}$.
\end{itemize}

Diagnostics are shown in \cref{fig:ablation-metrics} and the corresponding converged potential profiles in \cref{fig:ablation-profiles}.

\suppnote{Additional numerical details and potential profiles}
\label{sec:details}

\subsection*{Common numerical details}

Common hyperparameters of all calculations reported in the main text are collected in \cref{tab:hyperparams}.

\begin{table*}[t]
\centering
\footnotesize
\begin{tabular}{@{}l | p{2.1cm} p{2.9cm} p{2.4cm} p{1.9cm} p{3.0cm}@{}}
    \toprule
    & \textbf{Small systems} & \textbf{Dissociation} & \textbf{Rydberg} & \textbf{GW100} & \textbf{Larger systems} \\
    \midrule
    Orbital basis
    & \texttt{aug-cc-pVQZ}
    & \texttt{aug-cc-pwCV5Z}
    & \texttt{aug-cc-pVQZ} +diffuse
    & \texttt{aug-cc-pVTZ}
    & \texttt{cc-pVTZ} \\[2pt]
    $N_1$
    & $16 N_\mathrm{occ}$
    & $16 N_\mathrm{occ}$
    & $128$
    & $16 N_\mathrm{occ}$
    & $512$ (b), $1024$ (n) \\[2pt]
    $N_2$
    & $64 N_\mathrm{occ}$
    & $64 N_\mathrm{occ}$
    & $512$
    & $64 N_\mathrm{occ}$
    & $1024$ (b), $2048$ (n) \\[2pt]
    $\lambda$
    & $10^{-3}$
    & $10^{-3}$
    & $10^{-3}$
    & $10^{-3}$
    & $10^{-3}$ \\[2pt]
    Learning rate (peak)
    & $10^{-3}$
    & $10^{-3}$
    & $10^{-3}$
    & $10^{-3}$
    & $10^{-3}$ \\[2pt]
    Steps
    & $6000$
    & $6000$
    & $8000$
    & $4000$
    & $8000$ (b), $6000$ (n) \\[2pt]
    Spin
    & restricted
    & restricted
    & restricted
    & restricted
    & restricted \\
    Occupation
    & closed shell
    & closed shell; ROHF (ROKS H$_2^+$)
    & closed shell
    & closed shell
    & closed shell \\
    \bottomrule
\end{tabular}
\caption{
    \textbf{Common hyperparameters of the multipole-splat OEP and IKS calculations.}
    $N_1$ and $N_2$ are the numbers of monopole and dipole splats, $N_\mathrm{occ}$ is the number of occupied spatial orbitals of the reference, including singly occupied orbitals. Under "larger systems", benzene is tagged with (b) and naphthalene with (n) whenever values differ.
    ROHF and ROKS denote restricted open-shell Hartree-Fock and restricted open-shell Kohn-Sham, respectively.
}
\label{tab:hyperparams}
\end{table*}

Comments on individual hyperparameters:
\begin{itemize}
    \item \textbf{CCSD(T) reference}: For all IKS calculations, we use a relaxed CCSD(T) reference density computed in the same orbital basis and with the same restricted spatial-orbital convention as the inversion. The spin-summed one-particle reduced density matrix in the atomic-orbital basis is passed directly to the IKS optimization. All CCSD(T) calculations were carried out using PySCF~\cite{sunPySCFPythonbasedSimulations2018, sunRecentDevelopmentsPySCF2020}.
    \item \textbf{Regularization ($\lambda$):} The tabulated value is the exact-exchange value. Hybrid functionals (B3LYP, PBE0) scale $\lambda$ by the exact-exchange fraction $\gamma$ ($0.20$ and $0.25$). CCSD(T) inversion (IKS) uses the EXX $\gamma = 1$ value.
    \item \textbf{Optimizer and schedule:} The quoted learning rate is the peak of a cosine one-cycle schedule (Adamax, global gradient-norm clip $1.0$). An exponential moving average of the parameters with decay $0.99$ is applied except where noted.
    \item \textbf{Spin/occupation:} All calculations are spin-restricted, with a single shared spatial-orbital set and occupations in $\{0,1,2\}$.
\end{itemize}

Specific comments for individual systems and calculations:
\begin{enumerate}
    \item \textbf{Small systems}: Point-group symmetrization was used for atoms and symmetric linear molecules.
    \item \textbf{Dissociation:} Atom-Gaussian initialization was used instead of Boys localization, which is unreliable for one- and two-electron systems or stretched bonds. Grid level 5 and wide splat exponent bounds, $10^{-3} < \alpha < 10^{4}$, were used. The basis is \texttt{aug-cc-pwCV5Z} where available, with \texttt{aug-cc-pV5Z} used for the H and He curves. H$_2^+$ is a doublet ($N_\alpha-N_\beta=1$, or \texttt{spin=1} in PySCF), and its CCSD(T) inversion runs without density fitting.
    \item \textbf{Rydberg:} A custom diffuse Ne basis (\cref{tab:rydberg-basis}) was coupled with an extended radial grid (per-element radial scaling $\times 5$, grid level 5). An enlarged splat tube radius (tube scale $3.65$) and wide exponent bounds, $10^{-4} < \alpha < 10^{4}$, were used.
    \item \textbf{GW100}: The subset was restricted to molecules with $N \le 40$ electrons. Fewer optimization steps (4000) were used for affordability across the set.
    \item \textbf{Larger systems}: A non-augmented basis and reduced integration grid were used. The splats were initialized with $C_6$ symmetry for benzene and $D_{2h}$ symmetry for naphthalene.
\end{enumerate}

\subsection*{Reference spectral data processing}

Reference term values are taken from the Ne I level list in the NIST Atomic Spectra Database~\cite{kramidaNISTAtomicSpectra1999}. Ne I series converge to the two spin-orbit components of the Ne$^+$ core, $2s^2 2p^5\,{}^2P^\circ_{3/2}$ and $2s^2 2p^5\,{}^2P^\circ_{1/2}$. We use the lower ionization limit, $I_P = 0.7924823$ Ha, corresponding to $2s^2 2p^5\,{}^2P^\circ_{3/2}$. Each nominal $ns$ or $nd$ member splits into several fine-structure levels. We do not average these levels, but retain one consistently labeled branch of each series:
\begin{itemize}
	\item $ns$: $2s^2 2p^5({}^2P^\circ_{3/2})ns\,{}^2[3/2]^\circ_1$;
	\item $nd$: $2s^2 2p^5({}^2P^\circ_{3/2})nd\,{}^2[3/2]^\circ_1$.
\end{itemize}
These are the $J=1$, odd-parity branches in NIST's Racah notation and are dipole allowed from the ${}^1S_0$ ground state. All other fine-structure levels and the series converging to the upper ${}^2P^\circ_{1/2}$ core limit are excluded. NIST gives the term values $E_n$ relative to the Ne ground state. We convert them to the KS eigenvalue convention by
\begin{equation}
	\varepsilon_n^\text{NIST} = E_n^\text{NIST} - I_P,
\end{equation}
so that $\varepsilon_n^\text{NIST} < 0$ for every bound member and $\varepsilon_n^\text{NIST}\to0^-$ as $n \to \infty$, consistent with the sign convention used for our own OEP virtual eigenvalues. Quantum defects are extracted by inverting the Rydberg formula,
\begin{equation}
    \mu _l = n - \frac{1}{\sqrt{2 |\varepsilon _n ^\text{NIST} |}} \; ,
\end{equation}
evaluated separately for each series member. The defect remains member-dependent at low $n$. The values reported in the main text are obtained from an unweighted one-parameter least-squares fit of the Rydberg formula to the selected $n=3$--$6$ term values,
\begin{equation}
    \mu_l = \argmin_\mu \sum_{n=3}^{6} \left[ \varepsilon_n^\text{NIST} + \frac{1}{2(n-\mu)^2} \right]^2 \; .
\end{equation}
This gives $\mu_s=1.331$ and $\mu_d=0.012$. No $n\to\infty$ extrapolation or averaging over fine-structure levels is used.

The Ne Rydberg calculations use a customized orbital basis: the standard \texttt{aug-cc-pVQZ} set for neon, augmented with six additional single-primitive diffuse shells in each of the $s$, $p$, and $d$ angular momenta. Rydberg orbitals are hydrogenic and spatially extended. The $ns$ or $nd$ series up to $n \approx 6$ samples regions tens of Bohr from the nucleus, so they require Gaussians with exponents far smaller than those provided by valence-optimized sets. Each added series continues the geometric progression already present in the parent basis. Starting from the most diffuse exponent of that angular momentum in \texttt{aug-cc-pVQZ} ($0.1054$, $0.0818$, $0.2730$ a.u. for $s$, $p$, $d$, respectively), successive exponents are generated by repeated division by $3$ (six times), reaching $\sim 10^{-4}$ a.u. Continuing the existing even-tempered series keeps the added shells smoothly spaced and reduces the risk of near-linear dependence relative to an ad hoc overcomplete set. The extended basis is paired with an extended radial DFT grid (per-element radial scaling $\times 5$, grid level 5) and an enlarged splat tube radius (tube scale $3.65$), so that both the orbitals and the optimized effective potential have the spatial support needed to bind the Rydberg states. Numerical values of the additional exponents are listed in \cref{tab:rydberg-basis}.

\begin{table}[t]
\centering
\begin{tabular}{cccc}
    \toprule
    Shell & {$s$} & {$p$} & {$d$} \\
    \midrule
    1 & 0.035133 & 0.027260 & 0.091000 \\
    2 & 0.011711 & 0.009087 & 0.030333 \\
    3 & 0.003904 & 0.003029 & 0.010111 \\
    4 & 0.001301 & 0.001010 & 0.003370 \\
    5 & 0.000434 & 0.000337 & 0.001123 \\
    6 & 0.000145 & 0.000112 & 0.000374 \\
    \bottomrule
\end{tabular}
\caption{
    \textbf{Additional even-tempered diffuse basis exponents for the Ne Rydberg calculations.}
    Single primitives with unit contraction coefficient added to the aug-cc-pVQZ basis. Values given in a.u. Each column continues the parent basis's most diffuse exponent ($0.1054$, $0.0818$, $0.2730$ for $s$, $p$, $d$) by successive division by $3$.
}
\label{tab:rydberg-basis}
\end{table}

\subsection*{Additional potential profiles}

\begin{figure}[h]
    \centering
    \includegraphics[width=0.9\linewidth]{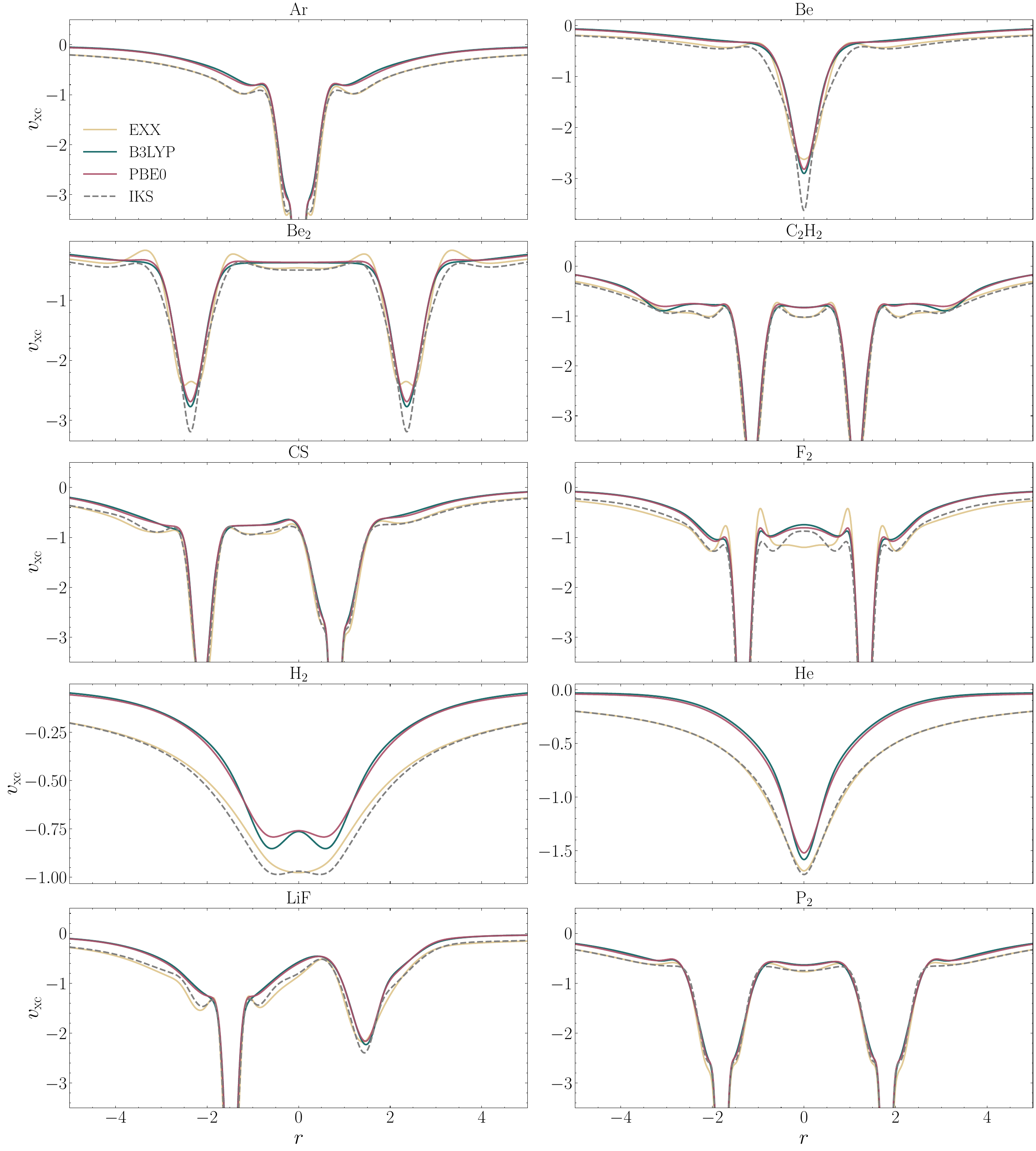}
    \caption{
        \textbf{Further examples of multipole splat potential spatial profiles.}
        A gallery of converged XC potentials $v_\xc$ for some of the remaining systems studied in this work, plotted along the internuclear axis and through the nucleus for atoms. EXX, B3LYP and PBE0 OEP solutions are contrasted with the IKS potential inverted from a CCSD(T) reference density.
    }
    \label{fig:potential-gallery}
\end{figure}

Localization energies and density total variations are reported in \cref{tab:energy-errs}. Corresponding converged potentials for the remaining systems studied in this work are collected in \cref{fig:potential-gallery}.

The B3LYP and PBE0 potentials in \cref{fig:potential-gallery} are nearly indistinguishable from one another throughout. Atomic and molecular shell structure is resolved as intershell shoulders and bond-region plateaus. The profiles show no visible numerical noise except for a peak inversion in Be${}_2$, which also requires special treatment in other OEP approaches~\cite{trushinNumericallyStableOptimized2021}.

\begin{table*}[h]
\centering
\begin{tabular}{cccc}
    \hline
    \multicolumn{4}{c}{Localization energy (mHa)} \\
    \toprule
    & EXX & B3LYP & PBE0 \\
    \midrule
    Ar & 1.091241 & 0.049067 & 0.063199 \\
    Be & 0.117724 & 0.015544 & 0.010047 \\
    Be$_2$ & 2.623287 & 0.133139 & 0.188917 \\
    C$_2$H$_2$ & 2.296297 & 0.106072 & 0.151531 \\
    CO & 2.931100 & 0.142226 & 0.199390 \\
    CS & 4.870364 & 0.241765 & 0.339502 \\
    F$_2$ & 5.739446 & 0.233667 & 0.364244 \\
    H$_2$ & 0.000000 & 0.009830 & 0.007424 \\
    He & 0.000000 & 0.005686 & 0.003209 \\
    LiF & 0.606390 & 0.057376 & 0.059881 \\
    N$_2$ & 2.892558 & 0.119665 & 0.180372 \\
    Ne & 0.239934 & 0.003968 & 0.004583 \\
    P$_2$ & 4.779071 & 0.232146 & 0.339860 \\
    SiO & 3.650403 & 0.190221 & 0.268817 \\
    \bottomrule
\end{tabular}
\hspace{1cm}
\begin{tabular}{cccc}
    \hline
    \multicolumn{4}{c}{Density total variation ($10^{-3}$ a.u.)} \\
    \toprule
     & EXX & B3LYP & PBE0 \\
    \midrule
    Ar & 0.631258 & 0.305103 & 0.284753 \\
    Be & 1.684929 & 0.827691 & 0.606656 \\
    Be$_2$ & 2.773294 & 0.665492 & 0.564080 \\
    C$_2$H$_2$ & 1.050024 & 0.361569 & 0.280723 \\
    CO & 1.103359 & 0.427190 & 0.400599 \\
    CS & 1.195568 & 0.466928 & 0.404159 \\
    F$_2$ & 0.727831 & 0.144382 & 0.155642 \\
    H$_2$ & 0.000026 & 1.253361 & 1.125515 \\
    He & 0.000026 & 0.571038 & 0.435304 \\
    LiF & 0.759181 & 0.361299 & 0.293866 \\
    N$_2$ & 0.928393 & 0.226618 & 0.219995 \\
    Ne & 0.477060 & 0.138863 & 0.130334 \\
    P$_2$ & 0.687671 & 0.261705 & 0.214088 \\
    SiO & 0.845790 & 0.344545 & 0.313913 \\
    \bottomrule
\end{tabular}
\caption{
    \textbf{Localization energies and density differences from generalized Kohn-Sham calculations.}
    \textbf{Left}: Localization energies $E_\text{loc} = E_\oep - E_\text{GKS}$ in mHa for exact exchange (EXX), B3LYP and PBE0 optimized effective potentials (OEP).
    \textbf{Right}: Density total variation $\mathcal D_\text{TV}$ between the converged OEP and the corresponding generalized Kohn-Sham (GKS) reference density.
}
\label{tab:energy-errs}
\end{table*}

\newpage
\bibliographystyle{naturemag}
\bibliography{references}